\documentclass{aa}
\usepackage{graphicx}                    
\usepackage{txfonts}

\newcommand{\mps}{m\,s$^{-1}$}
\newcommand{\bvec}[1]{ {\mathbf #1} }
\def\id{{\rm d}}

\def\cK{{\cal K}}
\def\cF{{\cal F}}

\def\cA{{\cal A}}
\def\vinv{v^{\rm inv}}
\def\ii{{\rm i}}

\begin{document}

\title{Estimate of the regularly gridded 3D vector flow field from a set of tomographic maps}

\author{Michal \v{S}vanda\inst{1,2}
\and
Marek Kozo\v{n}\inst{1}
}
\institute{Astronomical Institute, Charles University in Prague, Faculty of Mathematics and Physics, V Hole\v{s}ovi\v{c}k\'ach 2, CZ-18000 Prague 8, Czech Republic
\email{michal@astronomie.cz}
\and
Astronomical Institute (v. v. i.), Czech Academy of Sciences, Fri\v{c}ova 298, CZ-25165 Ond\v{r}ejov, Czech Republic
}

\abstract{
Time--distance inversions usually provide tomographic maps of the interesting plasma properties (we will focus on flows) at various depths. These maps however do not correspond directly to the flow field, but rather to the true flow field smoothed by the averaging kernels. We introduce a method to derive a regularly gridded estimate of the true velocity field from the set of tomographic maps. We aim mainly to reconstruct the flow on a uniform grid in the vertical domain. We derive the algorithm, implement it and validate using synthetic data. The use of the synthetic data allows us to investigate the influence of random noise and to develop the methodology to deal with it properly.}

\maketitle 

\section{SOLA helioseismic inversions in a nutshell}
The solar interior is filled with waves travelling between various places. In the convective envelope, the pressure (p) modes largely dominate the spectrum. The propagation of the waves is affected by perturbations in the plasma state parameters, by magnetic fields, and last but not least by plasma streaming. The time--distance helioseismology comprises a set of tools used to measure and analyse the wave travel times. In this study we focus on difference travel times, i.e. the difference of the measured travel time of waves travelling in the opposite directions. The difference travel times in the quiet Sun regions are mostly sensitive to flows \citep{2015SSRv..tmp...18B}.

The standard time--distance helioseismic pipeline consists of the following consecutive steps: first the spatio-temporal datacube is prepared using the tracking and mapping pipeline, this datacube is spatio-temporarily filtered to retain only waves of interest and subsequently the travel times are measured from cross-correlations of the filtered signal at two places. These travel times are finally inverted for flows assuming the linear relation between flow vector $\bvec{v}^{\rm true}$ and the measured difference travel time $\delta\tau$ via the sensitivity kernel $\bvec{K}^a$ \citep[coming usually from forward modelling, e.g.][]{2007AN....328..228B,2015SSRv..tmp...18B}

\begin{equation}
\delta\tau^a(\bvec{r})=\int_\odot \id^2\bvec{r'} \id z\, \bvec{K}^a (\bvec{r'}-\bvec{r},z) \cdot \bvec{v}^{\rm true}(\bvec{r'},z) + n^a(\bvec{r}),
\label{eq:forward}
\end{equation}
where the position vector splits into the horizontal $\bvec{r}$ and the vertical $z$ domains. Index $a$ uniquely refers to the selection of the measurement geometry (there is a free choice of spatio-temporal filters, distance of the measurement points, and/or additional spatial averaging). The realisation of the random noise $n^a$ is not known, however its covariance matrix may be measured from the data and used in the estimate of the random-noise level in the inverted flow velocities.  

Equation (\ref{eq:forward}) describes the forward problem which gives the recipe how to compute the (forward-modelled) travel times when the vector velocity field $\bvec{v}$ is known. The usual need is an inverse modelling, hence the derivation of the velocity field in the Sun from the measured set of travel-time maps. To do so, various classes of methods were employed, where two of them are used most often: the regularised least squares (RLS) and optimally localised averaging (OLA). The RLS method \citep[in time--distance helioseismology used for the first time by][]{1996ApJ...461L..55K} seeks to find the models of the solar interior, which provide the best least-squares fit to the measured travel-time maps, while regularising the solution (e.g., by requiring the smooth solution). OLA or its form Subtractive OLA \citep[SOLA;][]{1992AA...262L..33P} is based on explicitly constructed spatially confined averaging kernels by taking linear combination of sensitivity kernels, while simultaneously having control over the level of random noise in the results. A SOLA-type inversion is the principal method discussed in the current paper using a code validated by \cite{Svanda2011} and a data processing pipeline verified against the direct surface measurements by \cite{2013ApJ...771...32S}. 

Both methods result in the estimates of flows at a given target depth. This estimate is a true velocity field smoothed by the averaging kernel and it also contains the random-noise component ($v^{\rm rnd}$). 
\begin{equation}
\vinv_\alpha (\bvec{r}_0,z_0) = \int_\odot \sum\limits_\beta \cK^\alpha_\beta(\bvec{r}-\bvec{r}_0, z; z_0) v^{\rm true}_\beta(\bvec{r},z) \; \id^2\bvec{r}\,\id z + v_\alpha^{\rm rnd}(\bvec{r}_0,z_0),
\label{eq:inverse}
\end{equation}
where $\alpha=(x,y,z)$ is a $\alpha$-component of the velocity vector $\bvec{v}^{\rm true}$. SOLA method uses a user-given target function, which provides an initial estimate of the resulting averaging kernel. The method balances the spatial misfit of the true averaging kernel and the required target function and the propagation of the random noise into the resulting tomographic maps. 

The averaging kernel with components $\cK^\alpha_\beta$ can be derived from the inversion as a secondary product and it requires to be normalised so that its spatial integral equals to unity. The components contain information about the smoothing of the flow component in the direction of the inversion ($\beta=\alpha$) and also about the leakage (the cross-talk) from the other components (for $\beta\ne\alpha$). 

The level of the random noise may also be estimated directly from the inversion (in case of OLA method), so at least its root-mean-square (RMS) value $\sigma_\alpha={\rm RMS}(v_\alpha^{\rm rnd})$ is known. 

In effect, the output of SOLA inversion is a set of tomographic maps at various target depths. By the formulation, the results are therefore regularly gridded (defined on a regularly spaced grid) in the horizontal domain (where the spatial grid is usually regular), whereas in the vertical direction the coverage of the domain is given by a selection of the target depths, hence usually not regularly gridded. The regularly gridded solutions are required for instance whenever the estimates of spatial derivatives are needed, e.g. when computing for all components of divergence, vorticity, helicity, or Reynolds stress. The aim of this study is to derive and test the method to estimate the regularly gridded 3D vector flow field from such set of discrete tomographic maps with known averaging kernels and noise-level estimates. Simplified approaches were used recently, by usually using the target depths as true representative of the location of tomographic maps and interpolating between them in order to get an estimate of the regularly gridded velocity field \citep[e.g.][]{2015SoPh..290.1547D}. We would like to point out that such simplified methods are often not justified, because the typical helioseismic averaging kernels contain sidelobes in their sensitivity and then it is difficult to assign only a single depth to such a kernel. The averaging kernels also often peak at a different depth than where their e.g. gravity centre is located, which makes an assignment of a single representative depth even more ambiguous. Our goal is to find the way how to reconstruct a reasonable estimate of the vertical variations of the flow, which is only sparsely covered by a set of tomographic maps. We would like to use the full structure of the known averaging kernels to properly reconstruct the estimate of the true flow. Our technique aims to combine the information from a set of inversion with different averaging kernels, thereby to some extent resembling the super-resolution \citep[e.g.][]{Hardie1997} technique, routinely used in the imaging problems. 

\section{Formulation of the problem}
\label{sect:theproblem}
We will seek the 3D regularly gridded vector velocity field which, if convolved with a set of given averaging kernels, provides a set of flow maps that is close to the set of tomographic maps from the inversion at given target depths. Inspired by Eq.~(\ref{eq:inverse}) the sought relation may be written for each target depth $z^d$ as
\begin{equation}
\frac{1}{(\sigma^d_\alpha)^2} \vinv_\alpha(\bvec{r};z^d) = \frac{1}{(\sigma^d_\alpha)^2} h_x^2 h_z \sum\limits_{\beta,{\mathbf r}_j,z} \cK_\beta^\alpha(\bvec{r}_j-\bvec{r},z;z^d)v_\beta(\bvec{r}_j,z), 
\label{eq:oneline_real}
\end{equation}
where we replaced the continuous integrations by their discrete limits:
\begin{equation}
\int \id^2\bvec{r}\, f(\bvec{r}) \,\rightarrow \, \sum\limits_{{\mathbf r}_j} h_x^2 \,f(\bvec{r}_j) \enspace {\rm and} \enspace \int \id z\, f(\bvec{z}) \,\rightarrow \, \sum\limits_{z_j} h_z \,f(\bvec{z}_j).
\end{equation}
\begin{figure}[!t]
\centering
\includegraphics[width=0.43\textwidth]{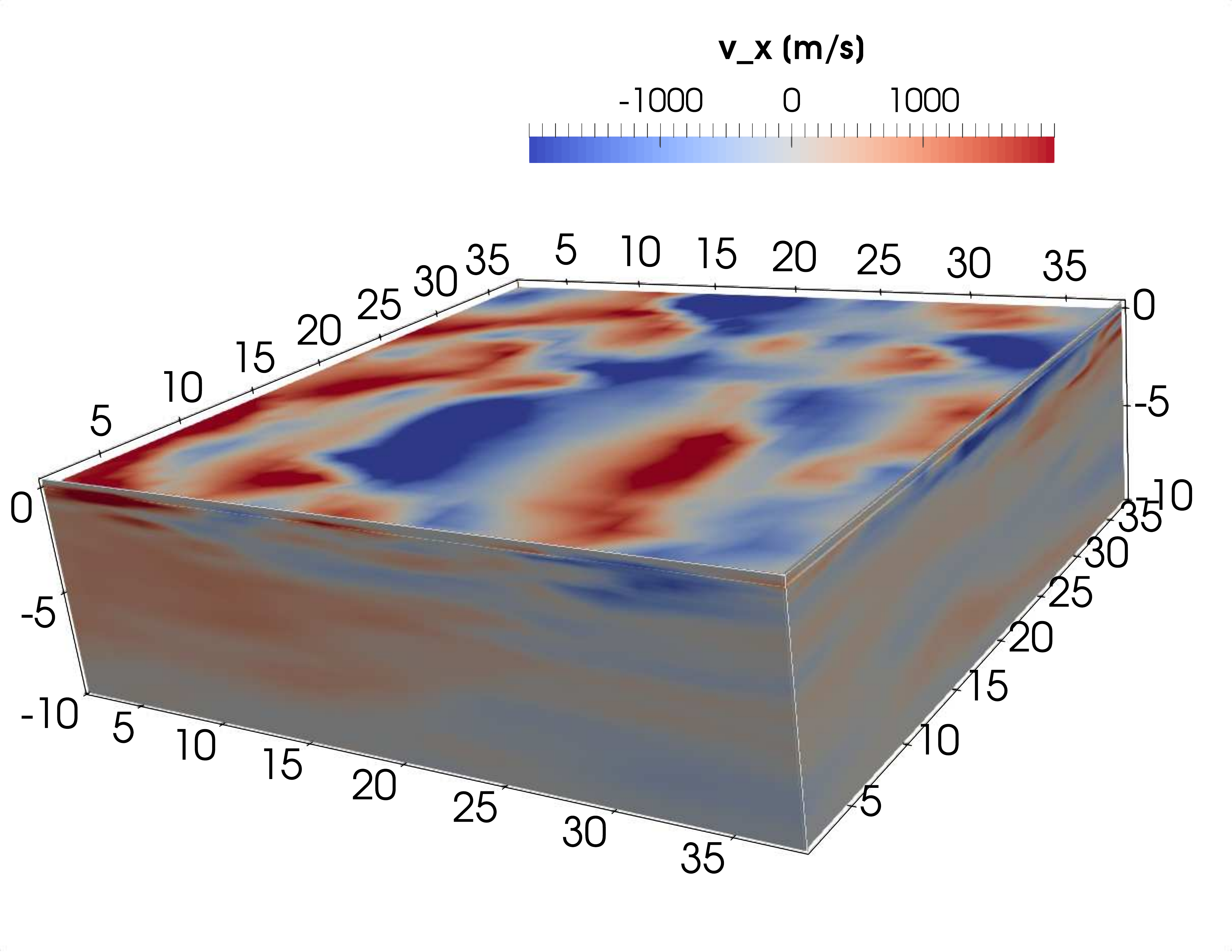}\\
\includegraphics[width=0.43\textwidth]{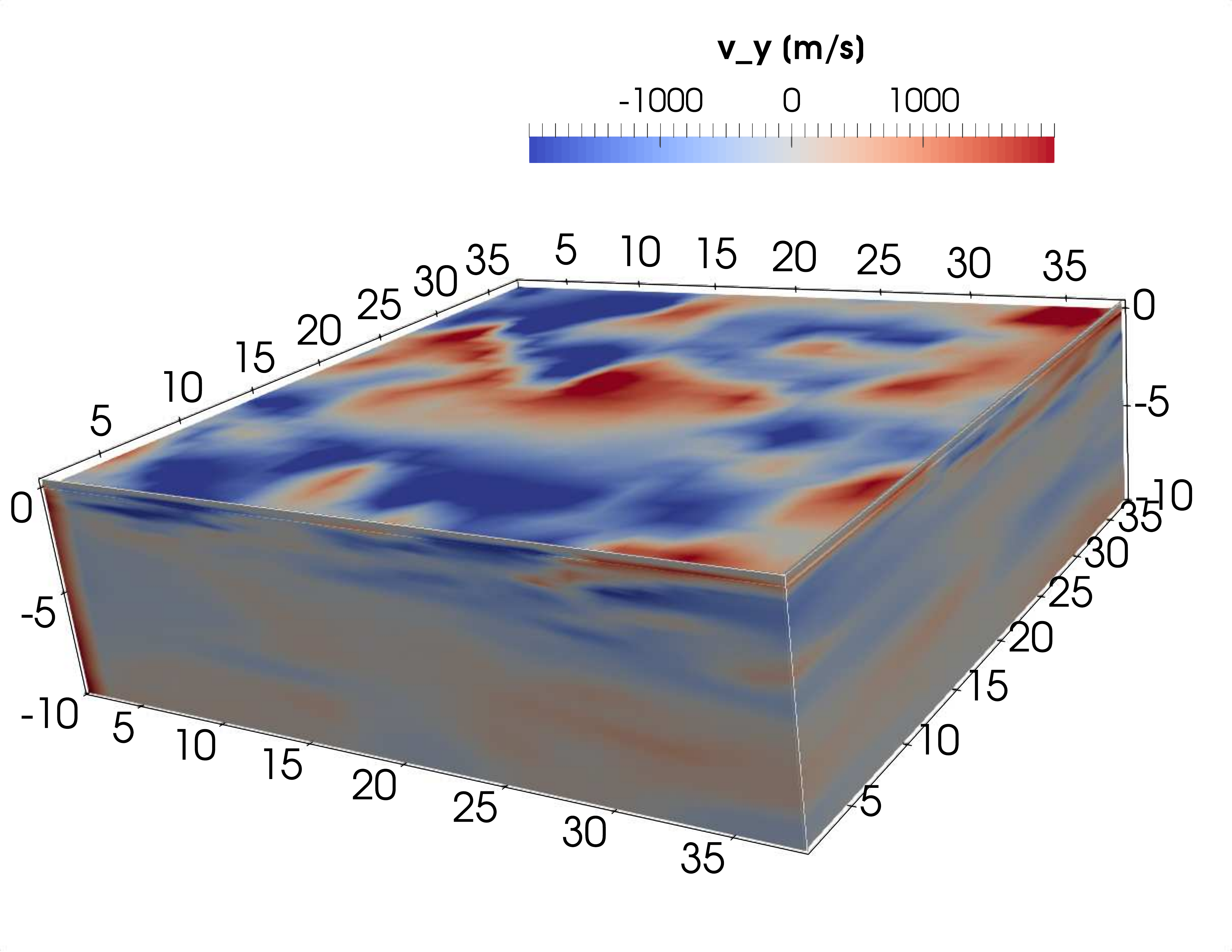}\\
\includegraphics[width=0.43\textwidth]{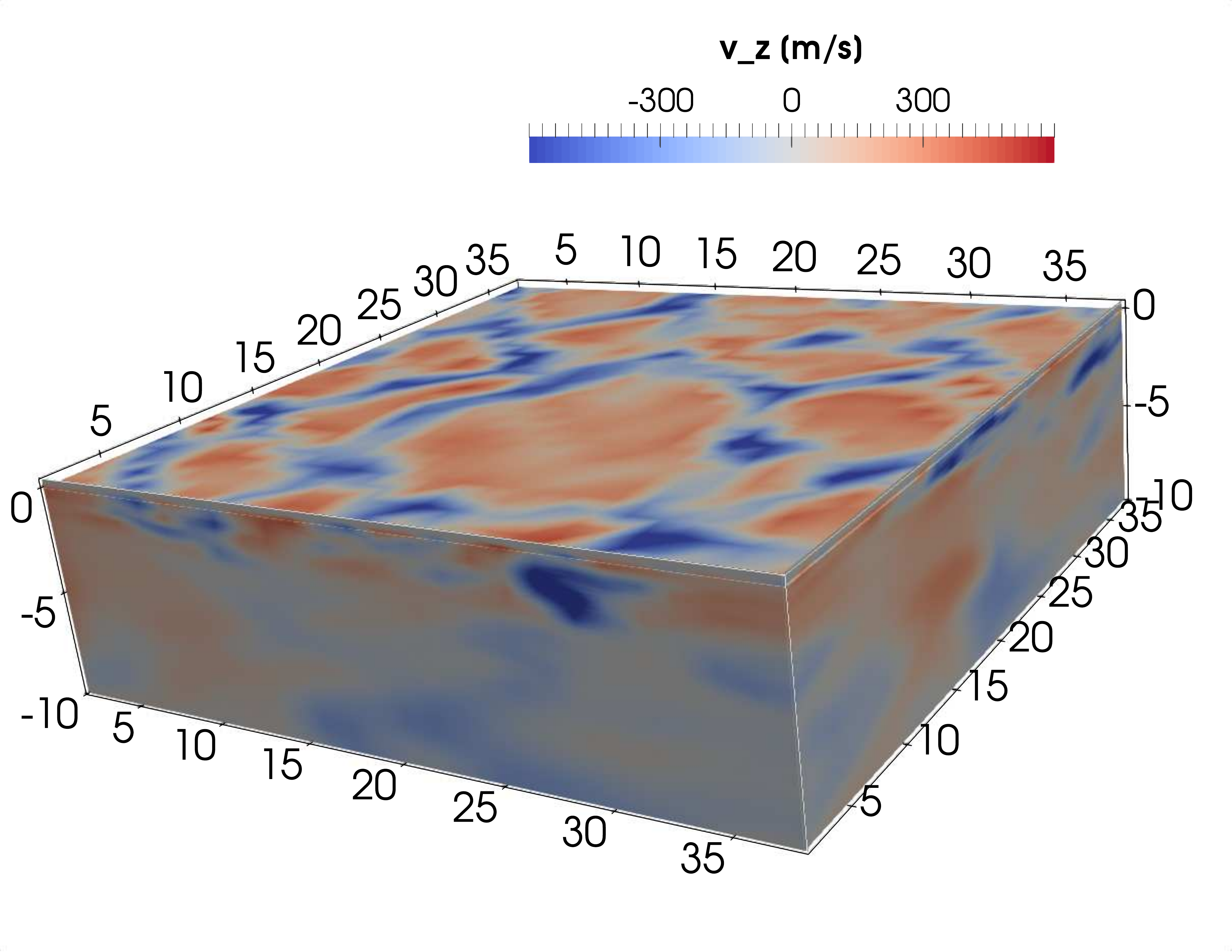}\\
\caption{Simulated convective flows used for the validation of our reconstruction techniques. The units of the coordinates are in Mm.}
\label{fig:synthetic}
\end{figure}
Note that Eq.~(\ref{eq:oneline_real}) is in fact a discretised Eq.~(\ref{eq:inverse}) without a noise term explicitly given. Constant $h_x$ is the spacing of the horizontal discrete grid (and $h_z$ is analogously the spacing in the vertical grid). For simplicity, we will assume that the grids in both domains are regular and equidistant. For irregular grids our solution must either be modified accordingly, or all vectors, matrices, and 3D arrays must be interpolated onto a regular grid. The symbolic summation over $\bvec{r}_j$ represents the summation over all $N=n_x n_y$ discrete values of position vector $\bvec{r}_j$ having dimensions $(n_x,n_y)$. Index $d$ indicates the selection of a target depth. We assume that the velocity vector $\bvec{v}$ is \emph{close} to the true velocity field $\bvec{v}^{\rm true}$. How close the two are will be discussed in Section~\ref{sect:akerns}. 

For each target depth $z^d$ there is a separate Eq.~(\ref{eq:oneline_real}), however the modelled velocity field $\bvec{v}$ is common to all these equations. The noise levels of each of these maps very likely vary, hence we introduced a inverse weighting by the RMS of the random noise, basically saying that less precise inferences (with larger RMS of random noise) are penalised against more precise inferences (with smaller RMS of random noise). For simplicity, we will first discuss the equations to be solved for one target depth and only then combine all available target depths to a bigger problem.

Factors of $1/(\sigma_\gamma^d)^2$ in (\ref{eq:oneline_real}) could in principle be removed from each side of the equation. In the following we will use a set of Equations (\ref{eq:oneline_real}) written for a set of target depths $z^d$, where keeping these factors in the equations will simplify an introduction of additional terms later, especially when a regularisation of the solution will be studied in order to bound the propagation of the random-noise.

Equation (\ref{eq:oneline_real}) may be written in a Fourier space with a definition of the Fourier transform given by a pair of equations
\begin{align}
\tilde{f}(\bvec{k},z)&= \frac{h_x^2}{(2\pi)^2}\sum_{\mathbf r} f(\bvec{r},z) \exp{\left[-\ii\bvec{k}\cdot \bvec{r}\right]},\label{eq:Ffor}\\
f(\bvec{r},z)&= h_k^2 \sum_{\mathbf k} \tilde f(\bvec{k},z) \exp{\left[ \ii\bvec{k}\cdot \bvec{r}\right]}\label{eq:Finv}, 
\end{align}
where $\bvec{k}$ is a horizontal wave vector and $h_k$ spacing in the Fourier domain:
\begin{equation}
\frac{1}{\left(\sigma^d_\gamma\right)^2} \tilde v_\gamma^{\rm inv}(\bvec{k};z^d) = \frac{(2\pi)^2}{\left(\sigma^d_\gamma\right)^2}   h_z  \sum_\beta\sum_z \tilde{\cK}^{\gamma*}_\beta(\bvec{k},z;z^d) \tilde v_\beta(\bvec{k},z),\quad \forall \bvec{k}.
\label{eq:oneline}
\end {equation}
The star $^{*}$ denotes the complex conjugate. Equation (\ref{eq:oneline}) holds for each wave vector $\bvec{k}$ and target depth $z^d$. The precise derivation of (\ref{eq:oneline}) is given in Appendix~\ref{app:oneline}. 

\subsection{Solution for the flow}
For each target depth $z^d$ we have three (three velocity components $v_\gamma$) Equations~(\ref{eq:oneline}). However, all these equations have a common velocity field $\tilde{\bvec{v}}(\bvec{k},z)$ that is to be determined. Not counting the ideal or trivial cases, in reality we expect that it will be impossible to fulfil Equations~(\ref{eq:oneline}) for a set of different target depths $z^d$ exactly. By casting the problem written for a set of target depths into the matrix equation
\begin{equation}
y(\bvec{k})=(2\pi)^2 h_z A(\bvec{k})x(\bvec{k}), \quad\forall \bvec{k}
\label{eq:yAx}
\end{equation}
the use of the pseudoinverse solvers will allow us to find the solution $x$ in a least-squares sense. Matrices $y$, $A$, and $x$ constitute of blocks of Eq.~(\ref{eq:oneline}), namely
\begin{equation}
y=
\begin{pmatrix}
        \begin{bmatrix}
                \frac{1}{\left(\sigma^{d_1}_x\right)^2} \tilde v_x^{\rm inv}(\mathbf k; z^{d_1})\\
                \frac{1}{\left(\sigma^{d_1}_y\right)^2} \tilde v_y^{\rm inv}(\mathbf k; z^{d_1})\\
                \frac{1}{\left(\sigma^{d_1}_z\right)^2} \tilde v_z^{\rm inv}(\mathbf k; z^{d_1})\\
        \end{bmatrix}\\
        \begin{bmatrix}
                \frac{1}{\left(\sigma^{d_2}_x\right)^2} \tilde v_x^{\rm inv}(\mathbf k; z^{d_2})\\
                \frac{1}{\left(\sigma^{d_2}_y\right)^2} \tilde v_y^{\rm inv}(\mathbf k; z^{d_2})\\
                \frac{1}{\left(\sigma^{d_2}_z\right)^2} \tilde v_z^{\rm inv}(\mathbf k; z^{d_2})\\
        \end{bmatrix}\\
        \vdots \downarrow (d)
\end{pmatrix},
\end{equation}
\begin{equation}
A=
\begin{pmatrix}
        \begin{bmatrix}
\frac{\tilde{\cK}^{x*}_x({\mathbf k},z;z^{d_1})}{\left(\sigma^{d_1}_x\right)^2}  & \frac{\tilde{\cK}^{x*}_y({\mathbf k},z;z^{d_1})}{\left(\sigma_x^{d_1}\right)^2} & \frac{\tilde{\cK}^{x*}_z({\mathbf k},z;z^{d_1})}{\left(\sigma^{d_1}_x\right)^2} \\
\frac{\tilde{\cK}^{y*}_x({\mathbf k},z;z^{d_1})}{\left(\sigma^{d_1}_y\right)^2}   & \frac{\tilde{\cK}^{y*}_y({\mathbf k},z;z^{d_1})}{\left(\sigma^{d_1}_y\right)^2}  & \frac{\tilde{\cK}^{y*}_z({\mathbf k},z;z^{d_1})}{\left(\sigma^{d_1}_y\right)^2}  \\
\frac{\tilde{\cK}^{z*}_x({\mathbf k},z;z^{d_1})}{\left(\sigma^{d_1}_z\right)^2}  & \frac{\tilde{\cK}^{z*}_y({\mathbf k},z;z^{d_1})}{\left(\sigma^{d_1}_z\right)^2} & \frac{\tilde{\cK}^{z*}_z({\mathbf k},z;z^{d_1})}{\left(\sigma^{d_1}_z\right)^2} 
                \end{bmatrix}&
\ldots \rightarrow (z)\\
        \begin{bmatrix}
\frac{\tilde{\cK}^{x*}_x({\mathbf k},z;z^{d_2})}{\left(\sigma^{d_2}_x\right)^2}  & \frac{\tilde{\cK}^{x*}_y({\mathbf k},z;z^{d_2})}{\left(\sigma_x^{d_2}\right)^2} & \frac{\tilde{\cK}^{x*}_z({\mathbf k},z;z^{d_2})}{\left(\sigma^{d_2}_x\right)^2} \\
\frac{\tilde{\cK}^{y*}_x({\mathbf k},z;z^{d_2})}{\left(\sigma^{d_2}_y\right)^2}   & \frac{\tilde{\cK}^{y*}_y({\mathbf k},z;z^{d_2})}{\left(\sigma^{d_2}_y\right)^2}  & \frac{\tilde{\cK}^{y*}_z({\mathbf k},z;z^{d_2})}{\left(\sigma^{d_2}_y\right)^2}  \\
\frac{\tilde{\cK}^{z*}_x({\mathbf k},z;z^{d_2})}{\left(\sigma^{d_2}_z\right)^2}  & \frac{\tilde{\cK}^{z*}_y({\mathbf k},z;z^{d_2})}{\left(\sigma^{d_2}_z\right)^2} & \frac{\tilde{\cK}^{z*}_z({\mathbf k},z;z^{d_2})}{\left(\sigma^{d_2}_z\right)^2} 
                \end{bmatrix}& \ldots \\
        \vdots \downarrow (d)
        \end{pmatrix},
\label{eq:A}
\end{equation}
and 
\begin{equation}
x=
        \begin{pmatrix}
                \begin{bmatrix}
                        \tilde v_x(\mathbf k,z)\\
                        \tilde v_y(\mathbf k,z)\\
                        \tilde v_z(\mathbf k,z)\\
                \end{bmatrix}\\
        \vdots \downarrow (z)
        \end{pmatrix}.
\end{equation}
The arrows in matrices $y$, $A$, and $b$ indicate the running indices in depth ($z$) and in target depths ($d$). The goal is to retrieve vector $x$, which contains the estimate of the regularly gridded velocity field. Such calculation may be done using standard solvers. Note that in case when $\cK_\alpha^\beta(\bvec{r},z;z^d)=0$ for $\alpha\ne\beta$ the solution will be simpler as the matrix $A$ will consist of diagonal blocks. Such solution is equivalent to the case when no cross-talks exist. Usually, there always is a (possibly small) contribution of the cross-talk so it is wise to use such information in the reconstruction of the estimate of the true velocity field. 
 
\subsection{Averaging kernels}
\label{sect:akerns}
In the previous section we assumed that the sought velocity field $\bvec{v}$ is close to the real flow $\bvec{v}^{\rm true}$. The relation ``close'' can be quantified by an assumed linear relation
\begin{equation}
\bvec{v}=\cF \bvec{v}^{\rm true} + \bvec{v}^{\rm noise},
\label{eq:akerns1}
\end{equation}
where $\bvec{v}^{\rm noise}$ corresponds to the propagated random noise and $\cF$ is a linear operator representing the convolution with the new averaging kernels $\cF^\alpha_\beta(\bvec{r},z;z_{\rm t})$. In this case $z_{\rm t}$ is the target depth on the vertical grid. For each grid point on the vertical grid we have a new averaging kernel. Equation (\ref{eq:akerns1}) is therefore an equivalent of Equation (\ref{eq:inverse}) in the original SOLA problem. Following the notation introduced in Eq.~(\ref{eq:akerns1}), (\ref{eq:inverse}) may also be written in a symbolic operator form
\begin{equation}
\bvec{v}^{\rm inv}= \cK \bvec{v}^{\rm true} + \bvec{v}^{\rm rnd}
\label{eq:akerns2}
\end{equation}
and (\ref{eq:yAx}) as
\begin{equation}
\bvec{v}= \cA^{-1} \bvec{v}^{\rm inv},
\label{eq:akerns3}
\end{equation}
where $\cA$ represents the operator form of matrix $A$ [Eq.~(\ref{eq:A})] that also absorbed the multiplicative factor of $(2\pi)^2 h_z$ for simplicity. 

We insert (\ref{eq:akerns1}) into (\ref{eq:akerns2}) and use (\ref{eq:akerns3}) to obtain
\begin{equation}
\cF \bvec{v}^{\rm true} + \bvec{v}^{\rm noise} = \cA^{-1} \cK \bvec{v}^{\rm true} + \cA^{-1} \bvec{v}^{\rm rnd}.
\end{equation}
By comparing the terms we have
\begin{equation}
\cF = \cA^{-1} \cK \quad {\rm and} \quad \bvec{v}^{\rm noise} = \cA^{-1} \bvec{v}^{\rm rnd}.
\label{eq:akernfinal}
\end{equation}

The equation above gives a recipe how to compute the new set of averaging kernels from the old ones and how does the realisation of the random noise propagate through the reconstruction. After doing the math in all details, the new averaging kernels $\cF^\alpha_\beta(\bvec{r},z;z_{\rm t})$ result from solving the equation
\begin{equation}
K^\alpha(\bvec{k})=(2\pi)^2 h_z A(\bvec{k})F^\alpha(\bvec{k}),\quad\forall \bvec{k}
\end{equation}
for each $\alpha$, $\bvec{k}$, and $z_t$, where
\begin{equation}
K^\alpha=
\begin{pmatrix}
        \begin{bmatrix}
                \frac{1}{\left(\sigma^{d_1}_x\right)^2} \tilde{\cK}^\alpha_x(\mathbf k,z; z^{d_1})\\
                \frac{1}{\left(\sigma^{d_1}_y\right)^2} \tilde{\cK}^\alpha_y(\mathbf k,z; z^{d_1})\\
                \frac{1}{\left(\sigma^{d_1}_z\right)^2} \tilde{\cK}^\alpha_z(\mathbf k,z; z^{d_1})\\
        \end{bmatrix}\\
        \begin{bmatrix}
                \frac{1}{\left(\sigma^{d_2}_x\right)^2} \tilde{\cK}^\alpha_x(\mathbf k,z; z^{d_2})\\
                \frac{1}{\left(\sigma^{d_2}_y\right)^2} \tilde{\cK}^\alpha_y(\mathbf k,z; z^{d_2})\\
                \frac{1}{\left(\sigma^{d_2}_z\right)^2} \tilde{\cK}^\alpha_z(\mathbf k,z; z^{d_2})\\
        \end{bmatrix}\\
        \vdots \downarrow (d)
\end{pmatrix},
\end{equation}

\begin{equation}
F^\alpha=
        \begin{pmatrix}
                \begin{bmatrix}
                        \tilde{\cF}^\alpha_x(\mathbf k,z;z_t)\\
                        \tilde{\cF}^\alpha_y(\mathbf k,z;z_t)\\
                        \tilde{\cF}^\alpha_z(\mathbf k,z;z_t)\\
                \end{bmatrix}\\
        \vdots \downarrow (z)
        \end{pmatrix}
\end{equation}
and $A$ is given by (\ref{eq:A}). 


\section{Random noise}
Tomographic maps from time--distance helioseismology naturally contain some level of random noise, which has the origin in the random excitation of the waves by convection. The realisation of this noise propagates through our reconstruction procedure into the regularly gridded estimate of the velocity field. 

\subsection{Random noise-level estimation}
\label{sect:random_noise}
Similarly to the determination of the averaging kernel we could proceed in deriving the propagation of random noise through the procedure, basically leading to the second part of Eq. (\ref{eq:akernfinal}). The tomographic maps as inputs to our reconstruction have known estimates of random noise-levels in the form of RMS of this random component only. As we will show later, this might not be enough as we will need (and use) also the estimate of the spatial power spectrum of this random-noise component $N(\bvec{k})$ (for details see Appendix~\ref{app:spectra}). In such case the use of the \emph{simulated} random-noise realisation seems to be a feasible solution to obtain an estimate of the random noise in the reconstructed flow cubes. 

Then the noise part of (\ref{eq:akernfinal}) written explicitly transforms into 
\begin{equation}
\delta y(\bvec{k})=(2\pi)^2 h_z A(\bvec{k})\delta x(\bvec{k}),\quad\forall \bvec{k}
\end{equation}
where $A$ is given by (\ref{eq:A}) and for $\delta x$ and $\delta y$ explicitly
\begin{equation}
\delta y=
\begin{pmatrix}
        \begin{bmatrix}
                \frac{1}{\left(\sigma^{d_1}_x\right)^2} \tilde v_x^{\rm rnd}(\mathbf k; z^{d_1})\\
                \frac{1}{\left(\sigma^{d_1}_y\right)^2} \tilde v_y^{\rm rnd}(\mathbf k; z^{d_1})\\
                \frac{1}{\left(\sigma^{d_1}_z\right)^2} \tilde v_z^{\rm rnd}(\mathbf k; z^{d_1})\\
        \end{bmatrix}\\
        \begin{bmatrix}
                \frac{1}{\left(\sigma^{d_2}_x\right)^2} \tilde v_x^{\rm rnd}(\mathbf k; z^{d_2})\\
                \frac{1}{\left(\sigma^{d_2}_y\right)^2} \tilde v_y^{\rm rnd}(\mathbf k; z^{d_2})\\
                \frac{1}{\left(\sigma^{d_2}_z\right)^2} \tilde v_z^{\rm rnd}(\mathbf k; z^{d_2})\\
        \end{bmatrix}\\
        \vdots \downarrow (d)
\end{pmatrix}, \quad {\rm and} \quad
\delta x=
        \begin{pmatrix}
                \begin{bmatrix}
                         \tilde v_x^{\rm noise}(\mathbf k,z)\\
                         \tilde v_y^{\rm noise}(\mathbf k,z)\\
                         \tilde v_z^{\rm noise}(\mathbf k,z)\\
                \end{bmatrix}\\
        \vdots \downarrow (z)
        \end{pmatrix}.
\end{equation}

Values of $v_\alpha^{\rm rnd}$ are generated randomly so that their spatial power spectrum is $N(\bvec{k})$ and their RMS equals to $\sigma_\alpha^{z_0}$. The estimate of the random-noise level $\sigma_\alpha^{z_{\rm t}}$ propagating to the resulting reconstructed velocity field is then computed as 
\begin{equation}
\sigma_\alpha^{z_{\rm t}}={\rm RMS}\left[v^{\rm noise}_\alpha(\bvec{r}; z_t)\right].
\end{equation}

\subsection{Dealing with random noise}
The algorithm described in Section~\ref{sect:theproblem} actually belongs to the problems of deconvolution in astrophysics, which are known to be unstable in presence of random noise. Using our synthetic tests we found that our implementation as described in the previous Sections in case of realistic noise power spectrum boosts the random noise by approximately eleven orders of magnitude, which is unacceptable for any reasonable flow reconstruction. 

Thus we need to find a method to suppress the propagation of the random noise. A way out is to avoid the spatial wave numbers, where the noise dominates. This can either be achieved by an introduction of the filter in the $\bvec{k}$-space domain that in effect allows to reconstruct only for some subset of wave vectors and to replace the contribution of wave vectors with large wave numbers (where the noise is likely do dominate) by zeros. We found that for a small level of noise (less than 1\% contribution) when around 20\% smallest wave numbers were used, the reconstruction led to an acceptable solution. For any larger noise contribution the solution was not acceptable. 

Another option is to introduce an ad-hoc regularisation term to ensure a smooth solution, by assuming that the random noise will cause small-scale oscillations in the solution. Such regularisation is quite common in helioseismology. We may introduce a smoothing parameter $\epsilon$ and thanks to keeping $1/(\sigma_\gamma^d)^2$ factors in Eq.~(\ref{eq:oneline}) simply modify the matrix $A$ to $A'$:
\begin{equation}
A'=A+
\begin{pmatrix}
        \begin{bmatrix}
\epsilon k^2  & 0 & 0 \\
0  &  \epsilon k^2 & 0\\
0 & 0 & \epsilon k^2 
                \end{bmatrix}&
\ldots \rightarrow (z)\\
        \vdots \downarrow (d)
        \end{pmatrix}.
\end{equation}

Our ad-hoc term resembles that of the RLS inversion as stated by e.g. \cite{2006ApJ...640..516C}. Then we invert for $x$ from equation $y=(2\pi)^2 h_z A'x$. The boosting of the noise by eleven orders mentioned above forces us to use a very strong regularisation (with $\epsilon$ in the order of $10^4$ or more), which in effect allows to reconstruct only very large-scale components of the spatial spectrum correctly, making the results extremely smooth. All the details of the flow field are lost. We do not find such result satisfactory. However, the regularisation may become handy e.g. when dealing with numerical stability, for instance when the matrix $A$ is close to singular.

As a final attempt we adapt solutions from the image restoration problems, namely the Wiener filtering. Wiener filtering introduces something like a $k$-space filter that allows to weight wave numbers so that only the wave numbers where the signal dominates the random noise contribute to the reconstructed image. The method requires prior knowledge about the expected power spectrum of the real solar flows and random noise, which we have at our disposal (details in Appendix~\ref{app:spectra}). 

Inspired by the Wiener-filtering algorithm instead of inverting for $x$ from $y=(2\pi)^2 h_z A x$ for each $\bvec{k}$ we obtain an optimal estimate $\hat{x}$ of $x$ as
\begin{equation}
\hat{x}=\frac{1}{(2\pi)^2 h_z} Gy,\quad {\rm where} \quad 
G(\bvec{k})=A^{-1}(\bvec{k}) \left[\frac{||A(\mathbf k)|| h_k^2}{||A(\mathbf k)||h_k^2+\frac{N(\mathbf k)}{S(\mathbf k)}} \right].
\end{equation}
Note that $A(\bvec{k})$ is a matrix, whereas in the original Wiener formulation it was a scalar. We represent the norm of the matrix by the Euclidean norm. $N(\bvec{k})$ and $S(\bvec{k})$ represent the estimates of power spectra of the random noise and the signal (for details see Appendix~\ref{app:spectra}).

\section{Implementation and synthetic tests}
\label{sect:tests}
\begin{figure}[!h]
\includegraphics[width=0.45\textwidth]{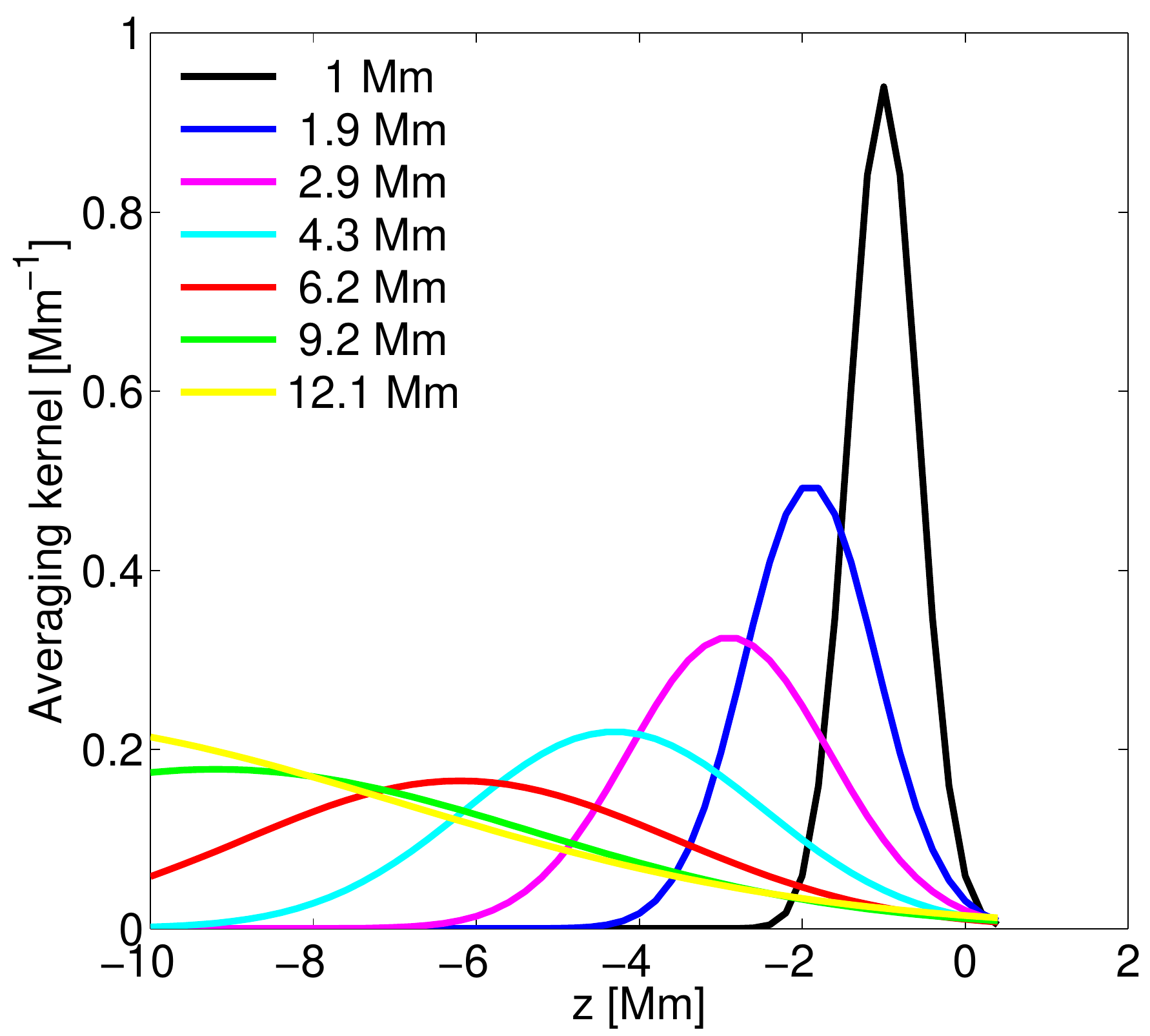}
\caption{Depth sensitivity of averaging kernels used in construction of seven synthetic tomographic maps.}
\label{fig:akerns}
\end{figure}

We implemented the above given algorithm in {\sc Matlab} programming language. To assess the performance of the method we constructed synthetic inversion-like tomographic maps which were obtained by smoothing the numerical simulation of the solar convection \citep[Fig.~\ref{fig:synthetic}, source data from ][]{2008ASPC..383...43U} with a set of averaging kernels (Fig.~\ref{fig:akerns}). The synthetic averaging kernels were constructed as Gaussians in both directions with a horizontal width of 5~Mm. In the vertical direction, the averaging kernels target depths of 1, 1.9, 2.9, 4.3, 6.2, 9.2, and 12.1~Mm with a full-width-at-half-maximum equal to the target depth of the given kernel. 

The reconstruction was performed on a horizontal grid identical to the horizontal grid of the model (with a pixel size of 1.4~Mm) and the vertical grid extending from 10~Mm depth to 400~km above the surface with a vertical sampling of 200~km. For each grid point we computed the estimate of the reconstructed flow vector, averaging kernels for each flow component, and estimate of the noise levels also for each flow component. Due to the assumed translation invariance in the horizontal domain, for each flow component the averaging kernels and the estimates of the noise levels are identical for all horizontal points within the identical depth. 

\begin{figure}
\includegraphics[width=0.45\textwidth]{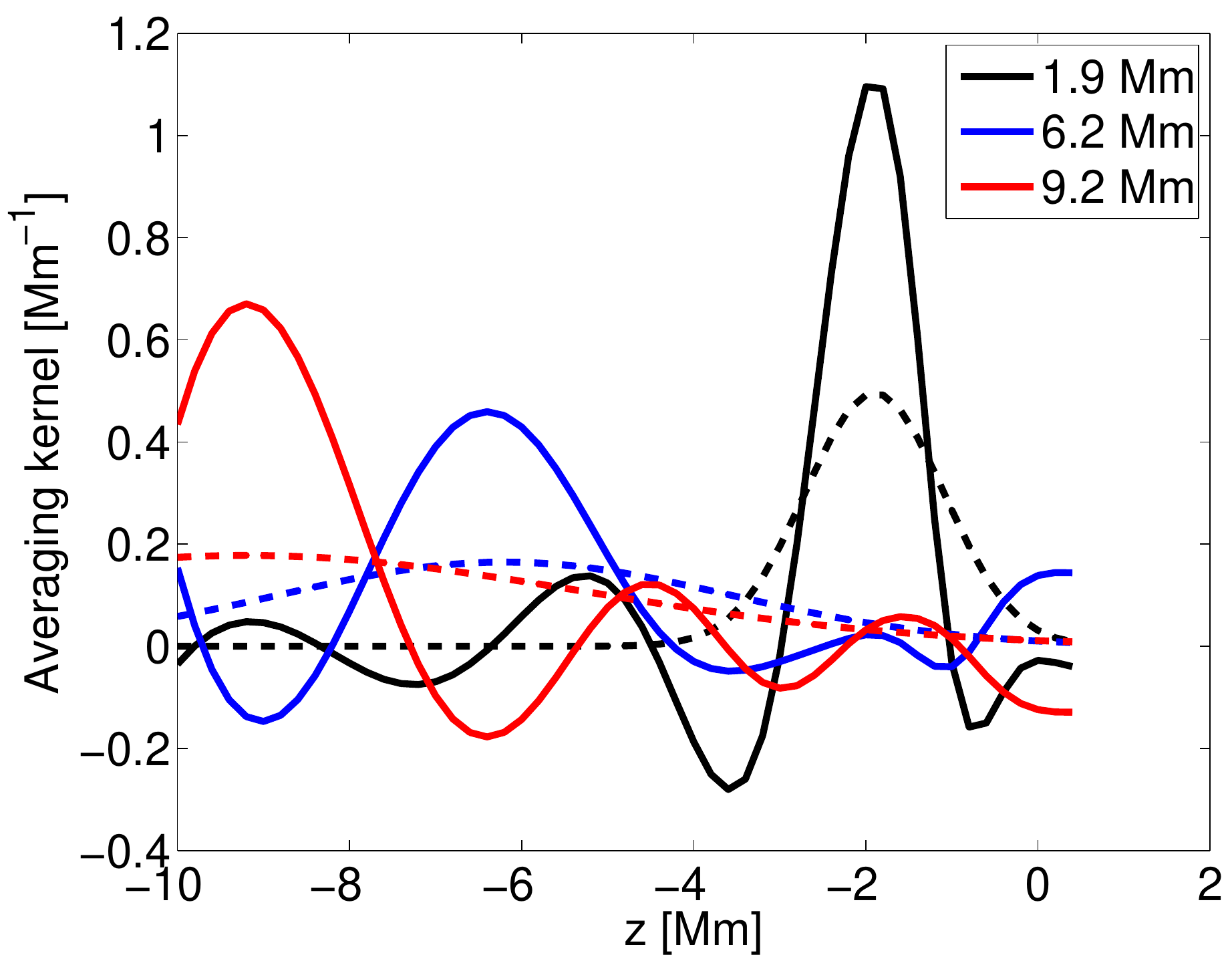}
\caption{Examples of original averaging kernels (dashed) and averaging kernels after deconvolution at the same depths (solid). Only horizontally averaged kernels are plotted. }
\label{fig:comparison_akerns}
\end{figure}

\begin{figure}
\includegraphics[width=0.45\textwidth]{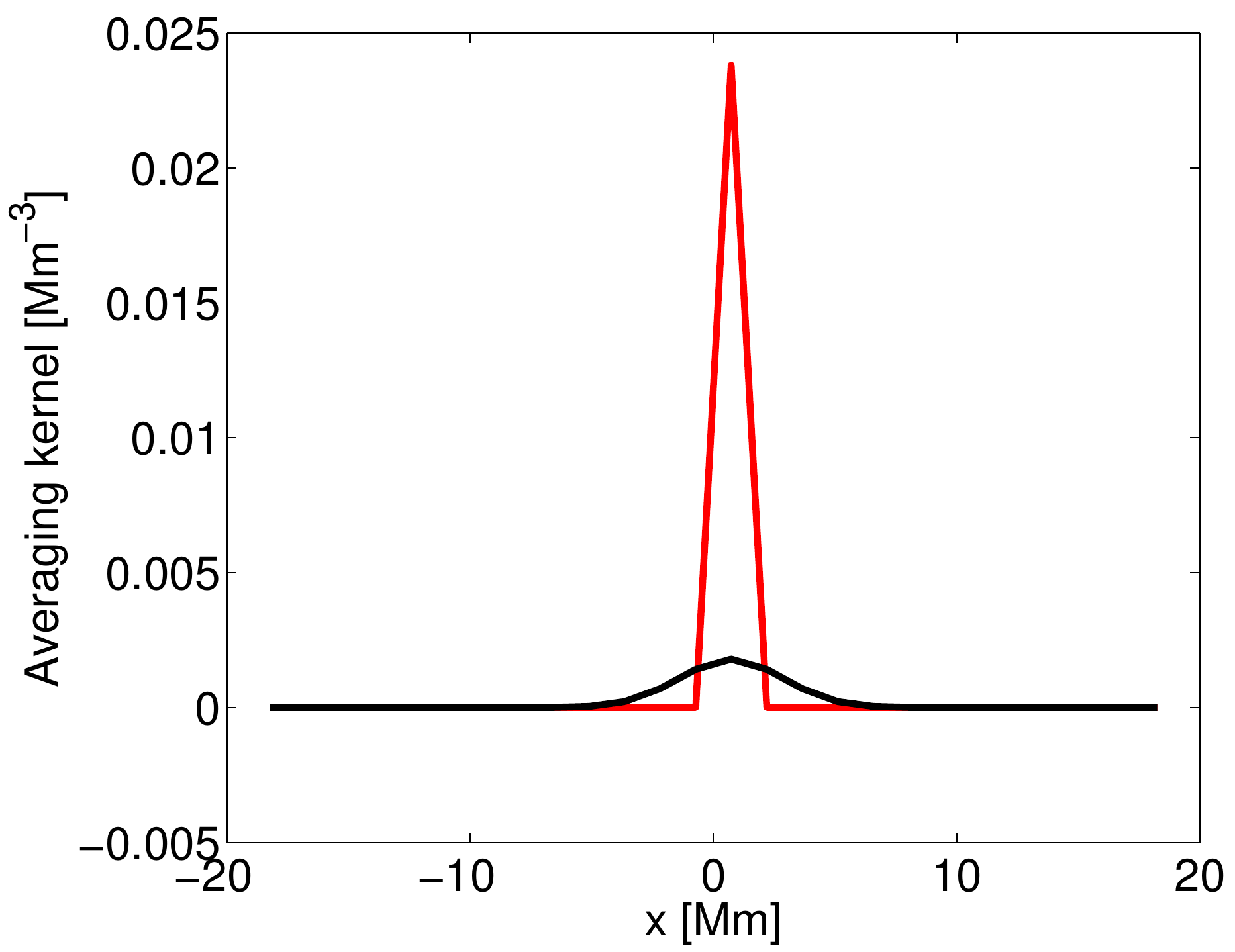}
\caption{Comparison of horizontal cuts through averaging kernels (along $y=0$) at the target depth 1.9~Mm for the input averaging kernel (black) and the reconstructed one (red). }
\label{fig:comparison_akerns_h}
\end{figure}

\begin{figure*}[!t]
\raisebox{5cm}{{\sf a)}}\includegraphics[width=0.49\textwidth]{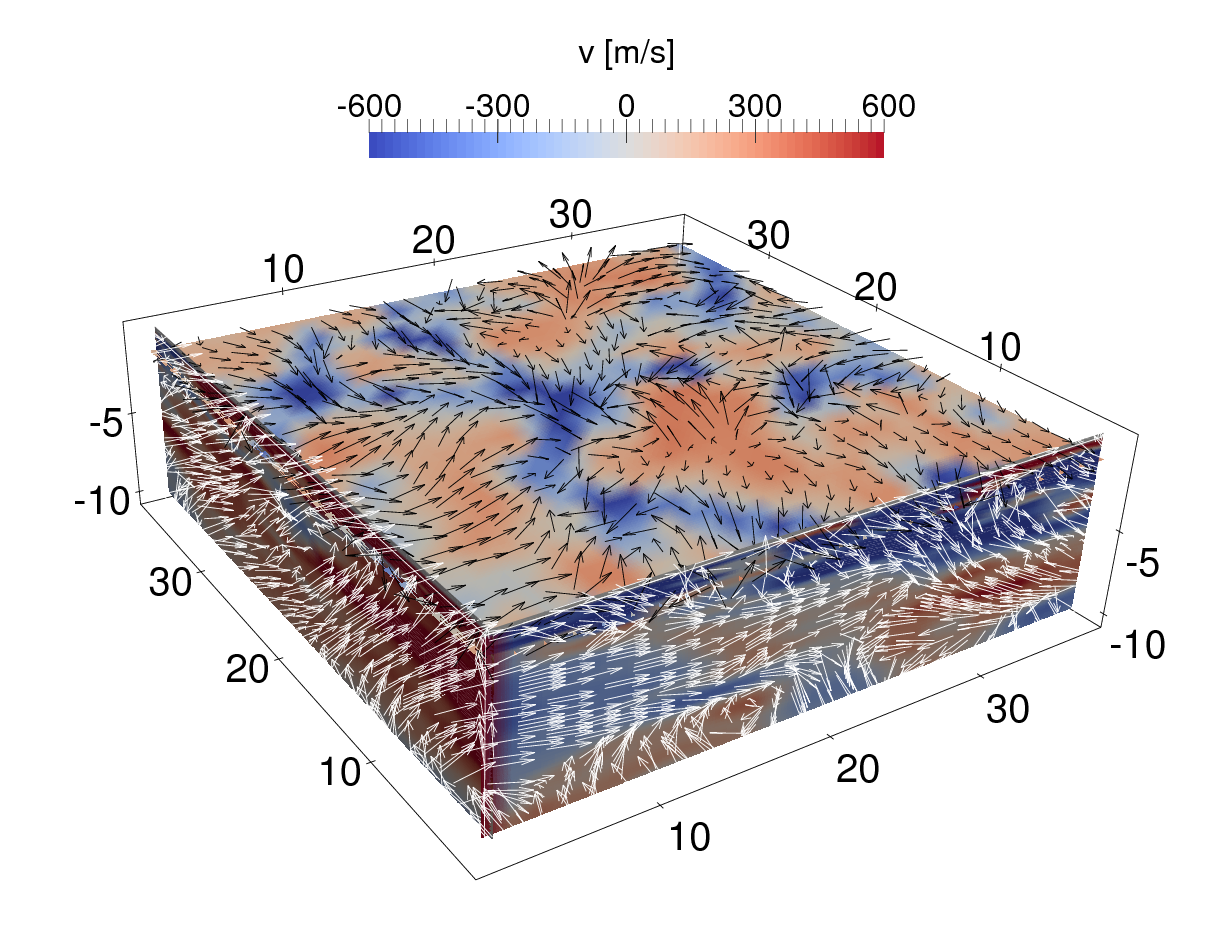}
\raisebox{5cm}{{\sf b)}}\includegraphics[width=0.49\textwidth]{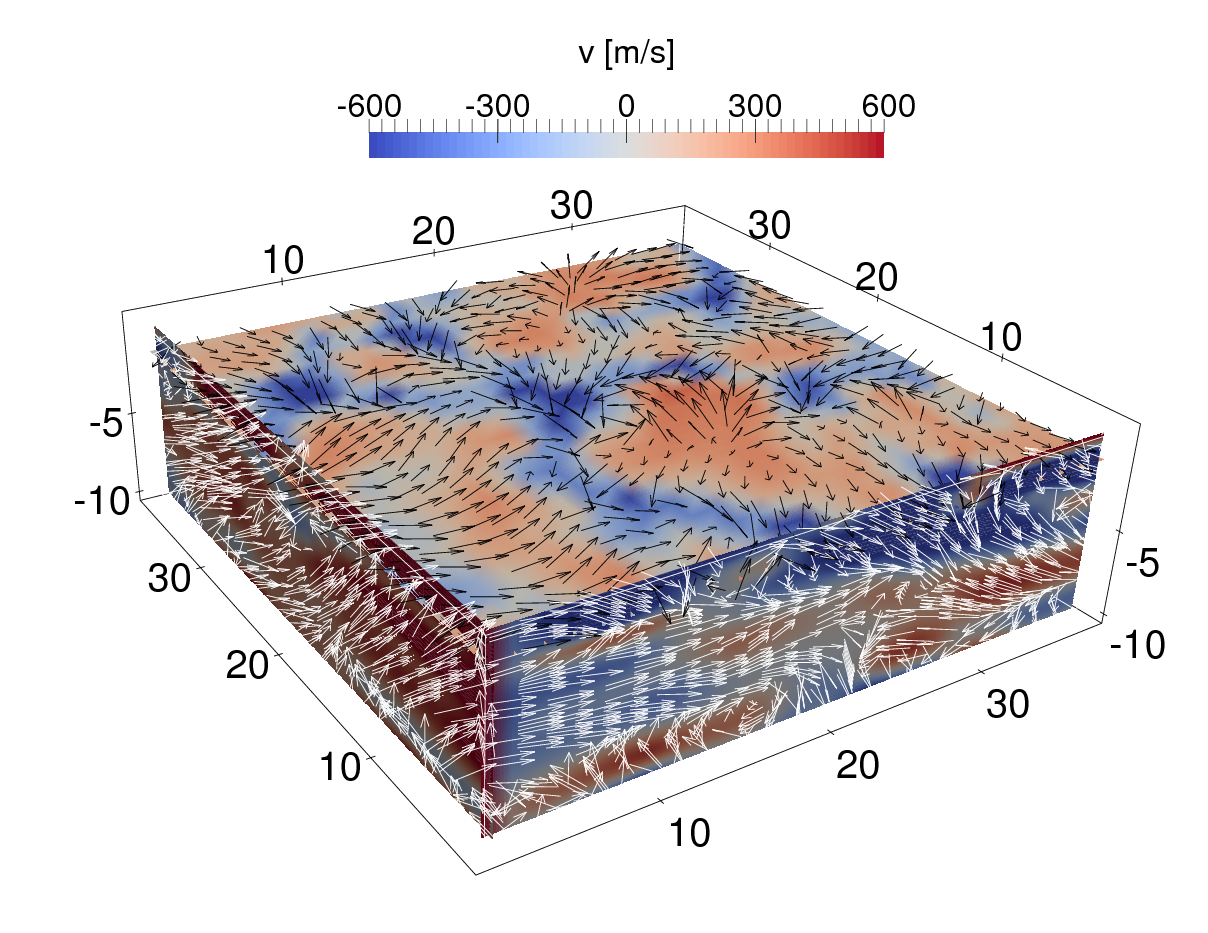}\\
\raisebox{5cm}{{\sf c)}}\includegraphics[width=0.49\textwidth]{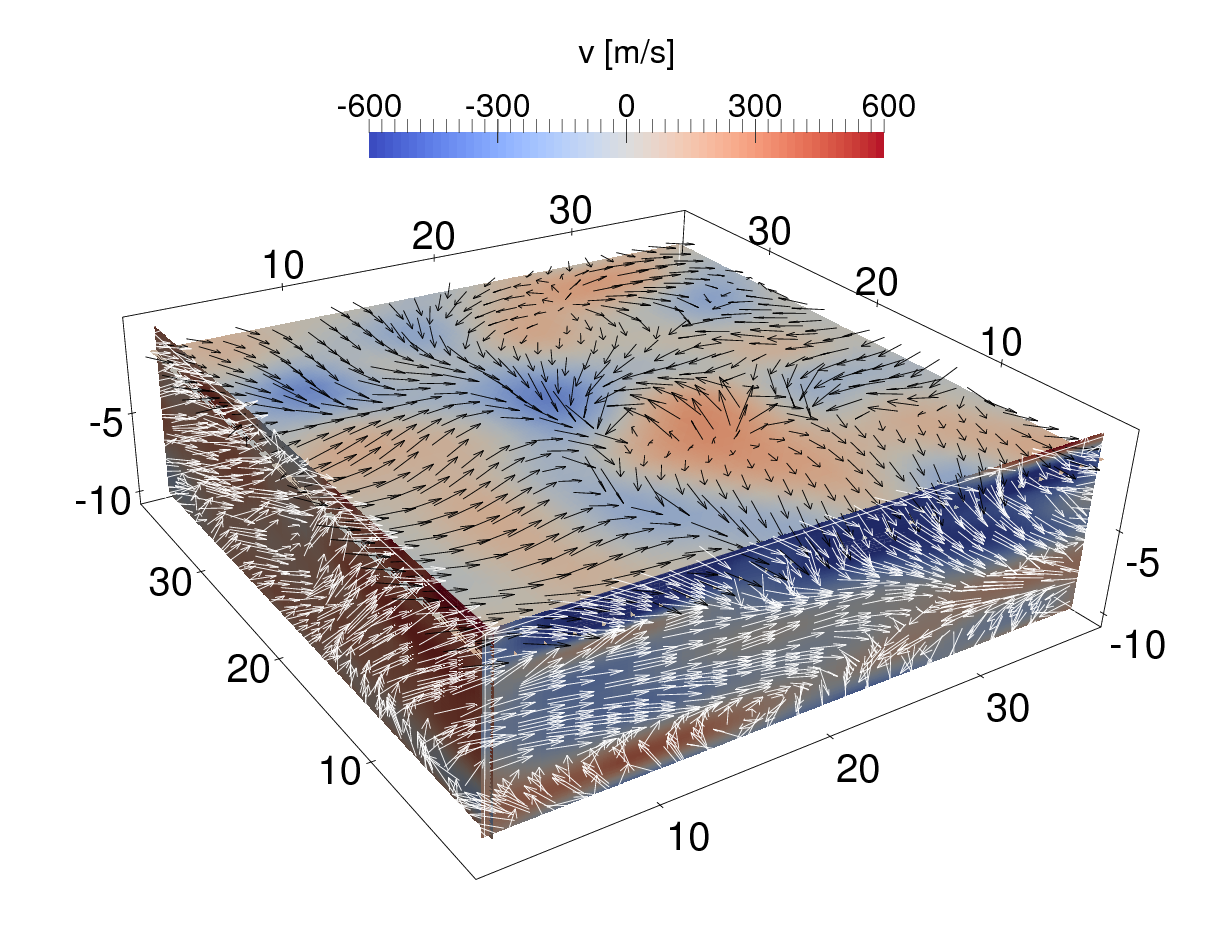}
\raisebox{5cm}{{\sf d)}}\includegraphics[width=0.49\textwidth]{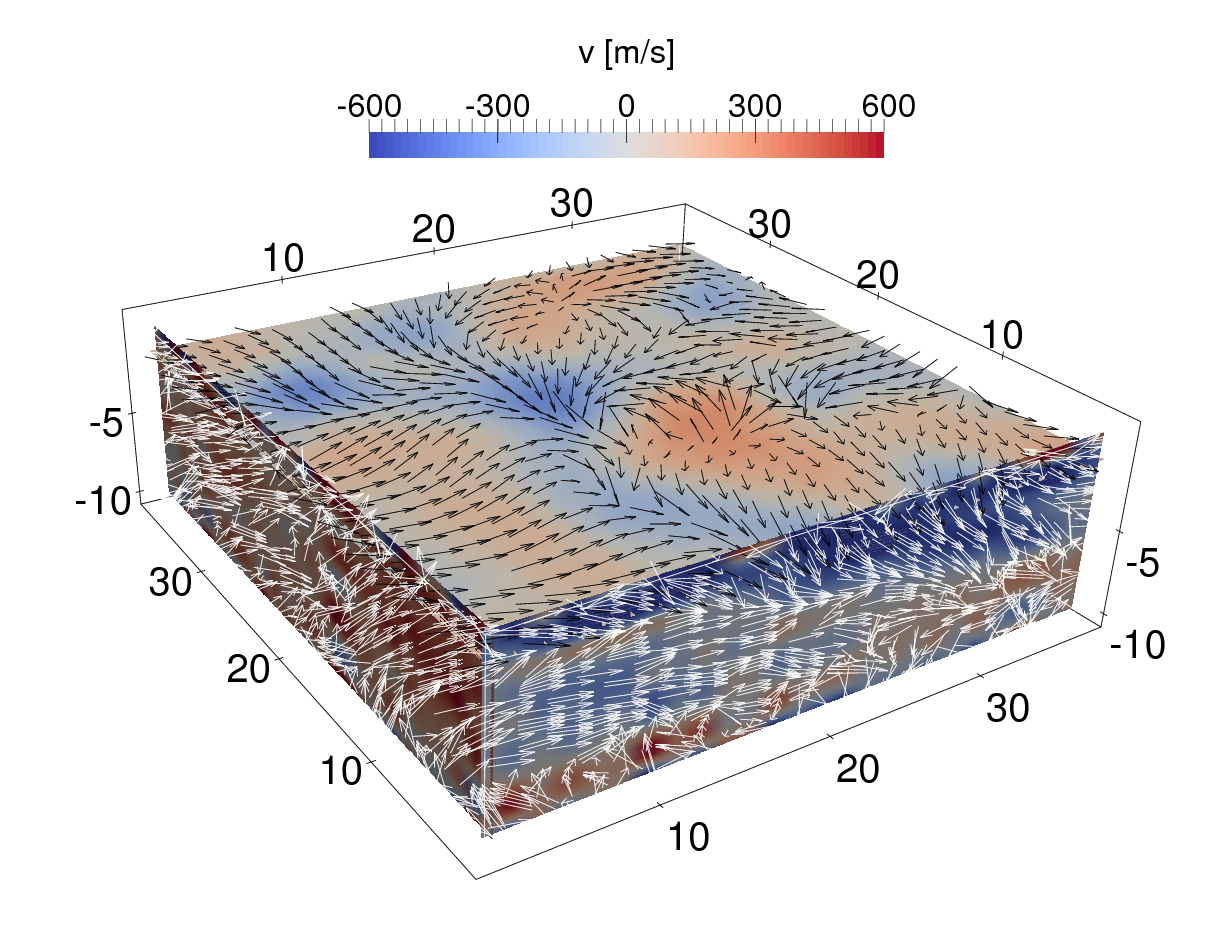}\\
\raisebox{5cm}{{\sf e)}}\includegraphics[width=0.49\textwidth]{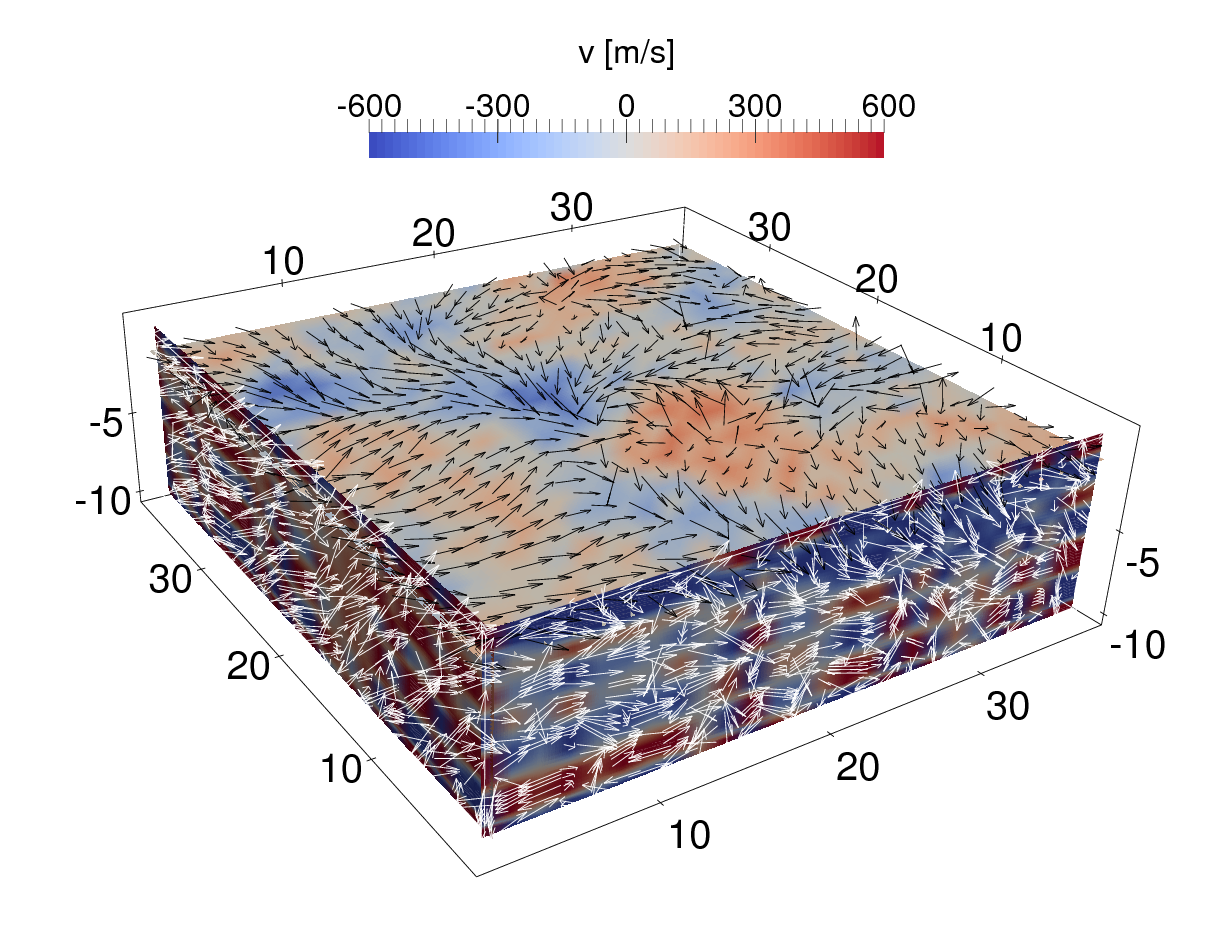}
\raisebox{5cm}{{\sf f)}}\includegraphics[width=0.49\textwidth]{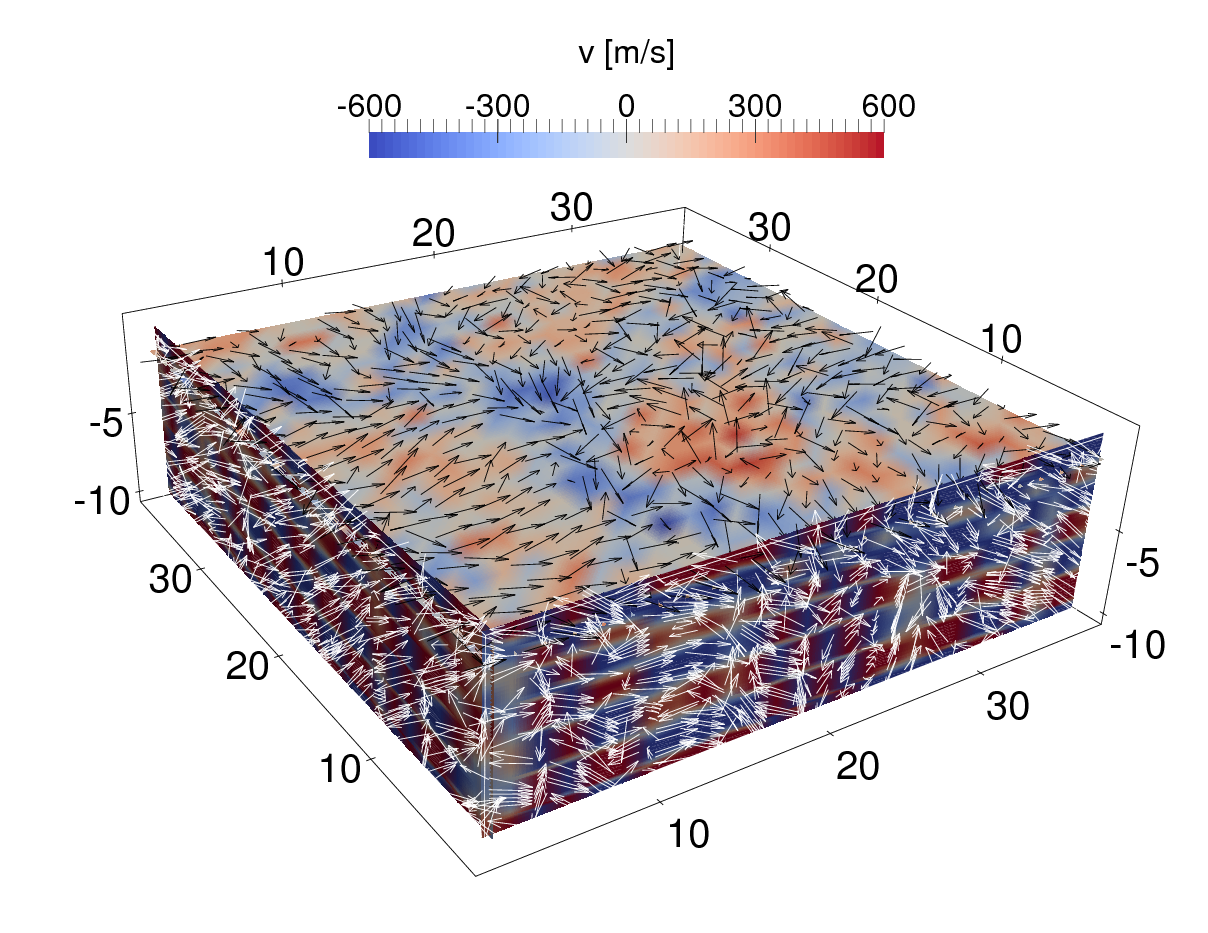}
\caption{Effect of the random noise to the reconstruction of the regularly gridded velocity field: the case of the near-surface flows. Displayed are results with varying noise level: (a) input data, (b) no noise. Other maps were reconstructed with Wiener-like filtering to deal with random-noise boosting with varying SNR levels: (c) SNR 1000, (d) SNR 100, (e) SNR 20, and (f) SNR 10. In each cut the parallel components are indicated by arrows, while the perpendicular component is indicated by the colour. Colour scale of all panels are identical. The units of the coordinates are in Mm. The maps of all three components at the depth of 1~Mm are displayed in Fig.~\ref{fig:comparisonmaps}.}
\label{fig:comparison}
\end{figure*}

\begin{figure*}[!t]
\centering
\includegraphics[width=0.75\textwidth]{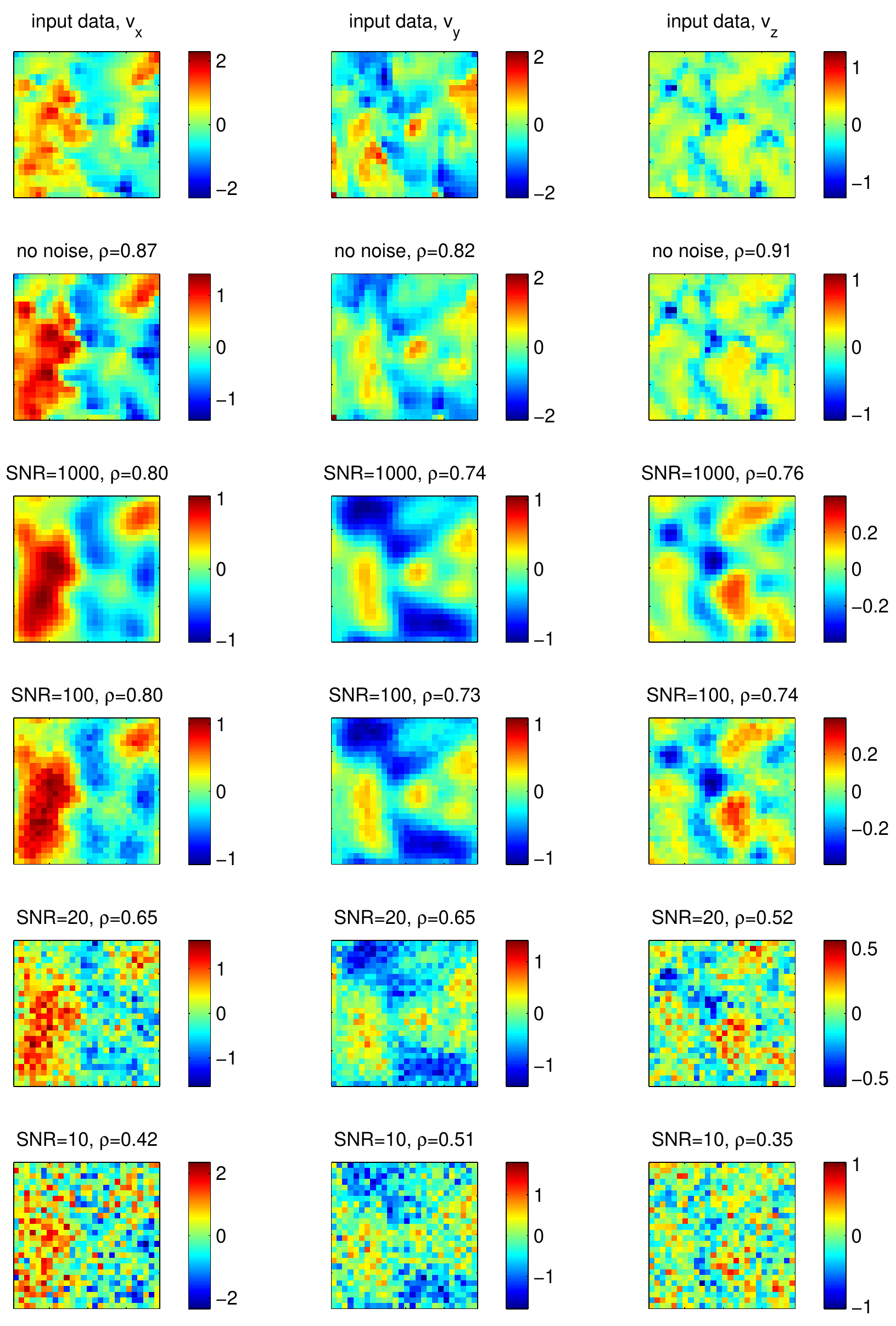}
\caption{Maps of all components of the flow at the depth of 1~Mm for various levels of the random noise. In the left-hand column one finds the maps for the $x$ component of velocity, in the middle one for the $y$ component, and in the right-hand one the maps for the vertical $z$ component. The rows contain the input velocity field in the first row and reconstructions with increasing level of random noise in the top-bottom order. Panels with given SNR value were reconstructed using Wiener-like approach. The correlation coefficient for the whole cube between the input velocity field and the reconstruction is given in the titles of the plot panels. Colour-bars have the units of k\mps{}, the field-of-view shown here covers 38~Mm\,$\times$\,38~Mm in the horizontal directions. }
\label{fig:comparisonmaps}
\end{figure*}

\begin{figure*}[!t]
\centering
\includegraphics[width=0.75\textwidth]{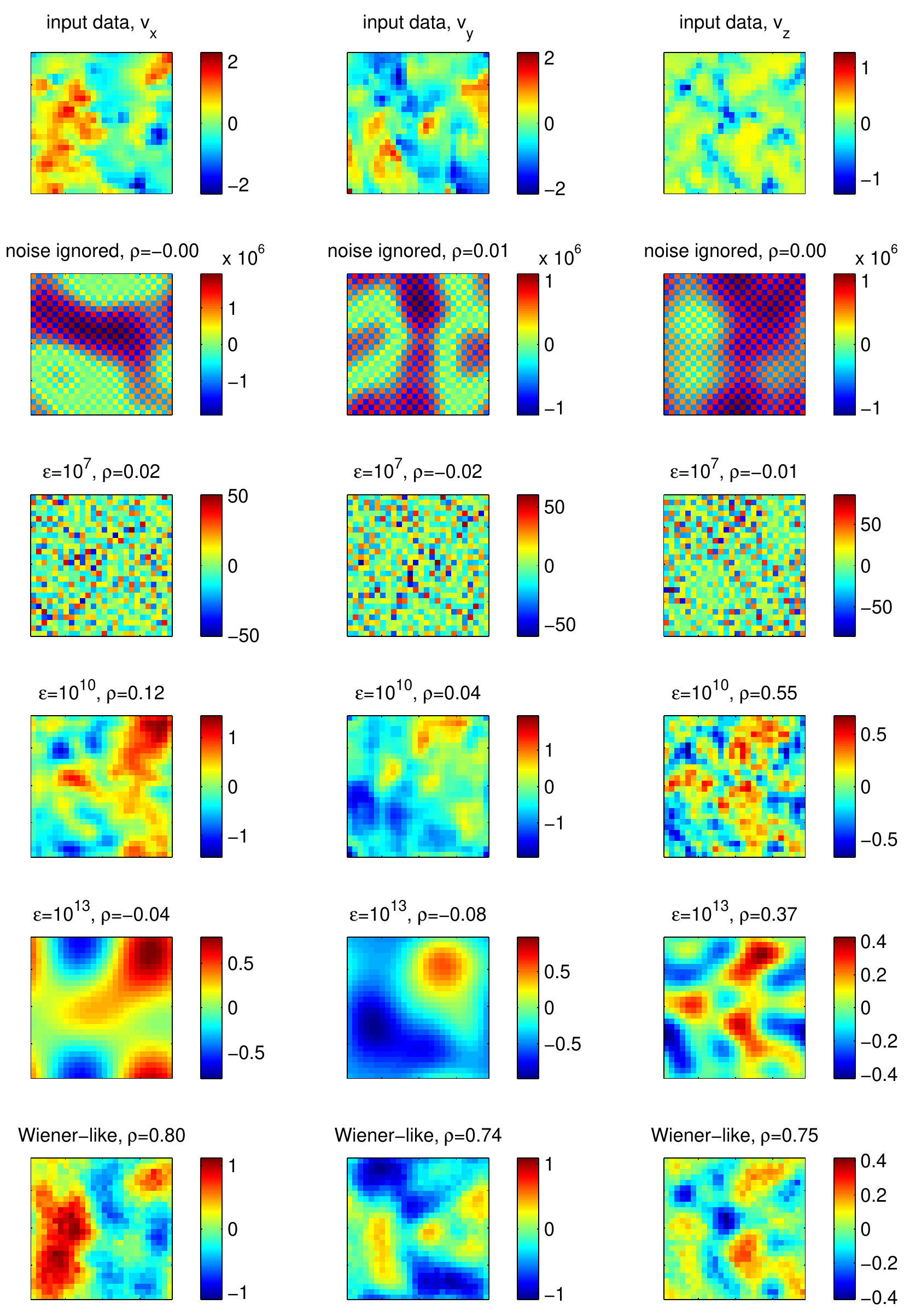}
\caption{Test of approaches to deal with a random noise. An input datacube having SNR of 100 (i.e. 1\% of random noise contribution -- the first row) was reconstructed with the basic method by ignoring the noise (second row), ad-hoc regularisation with a set of values of regularisation parameter (third to fifth row), and the Wiener-like filtering approach (last row). The correlation coefficient for the whole cube between the input velocity field and the reconstruction is given in the titles of the plot panels. Colour-bars have the units of k\mps{} (violet indicates fast variations from pixel to pixel), the field-of-view shown here covers 38~Mm\,$\times$\,38~Mm in the horizontal directions.}
\label{fig:comparisonmaps_methods}
\end{figure*}

The averaging kernels $\cF^\alpha_\beta$ are in general narrower than the averaging kernels $\cK^\alpha_\beta$ of the input set of tomographic maps (see Fig.~\ref{fig:comparison_akerns}), which is what we wanted. In Fig.~\ref{fig:comparison_akerns} we compared only averaging kernels at corresponding target depth, we remind the reader that due to the sampling of the vertical domain as described above, there is in fact in total 53 averaging kernels, one for each grid point in the vertical domain. In our testing case the overall shape of the new averaging kernels $\cF$ is not satisfying compared to the original ones, as the averaging kernels of the reconstruction contain many sidelobes, also the negative ones. It has to be noted that in a realistic case the averaging kernels $\cK$ also contain sidelobes (the negative ones as well), therefore from this point of view it is not possible to immediately say which is ``worse''.  

In Fig.~\ref{fig:comparison_akerns} we plotted examples of horizontally integrated kernels for simplicity. The horizontally integrated kernels are independent of noise levels, as by using the Wiener-like filtering the $\bvec{k}=0$ is always reconstructed. In the horizontal domain the shape of the kernels is influenced by the level of random noise. In the ideal case (no noise present) the deconvolution goes basically to the level of a pixel size (see Fig.~\ref{fig:comparison_akerns_h}), they are close to a numerical Dirac $\delta$ function. The described behaviour is exactly the same for all depths as the input averaging kernels had the same full-width-at-half-maximum of 5~Mm. With an increasing level of noise the kernels get wider and they approach the shape of the input kernels for larger noise levels. Then, in effect, only a little or no deconvolution of the flow in the horizontal domain takes place. We would like to stress that our main goal was to reconstruct the reasonable estimate of the flow in the vertical domain, where the input set of tomographic maps covers the vertical domain only sparsely.

The total integrals of the new averaging kernels are then used to properly scale the magnitude of the flows at each target depth. This re-normalisation is necessary because Wiener-like filtering suppresses some of the Fourier components and therefore naturally introduces a scaling of resulting flow magnitudes. Consequently, we normalise new averaging kernels so that their spatial integral equals to unity. 

\begin{figure}
\includegraphics[width=0.45\textwidth]{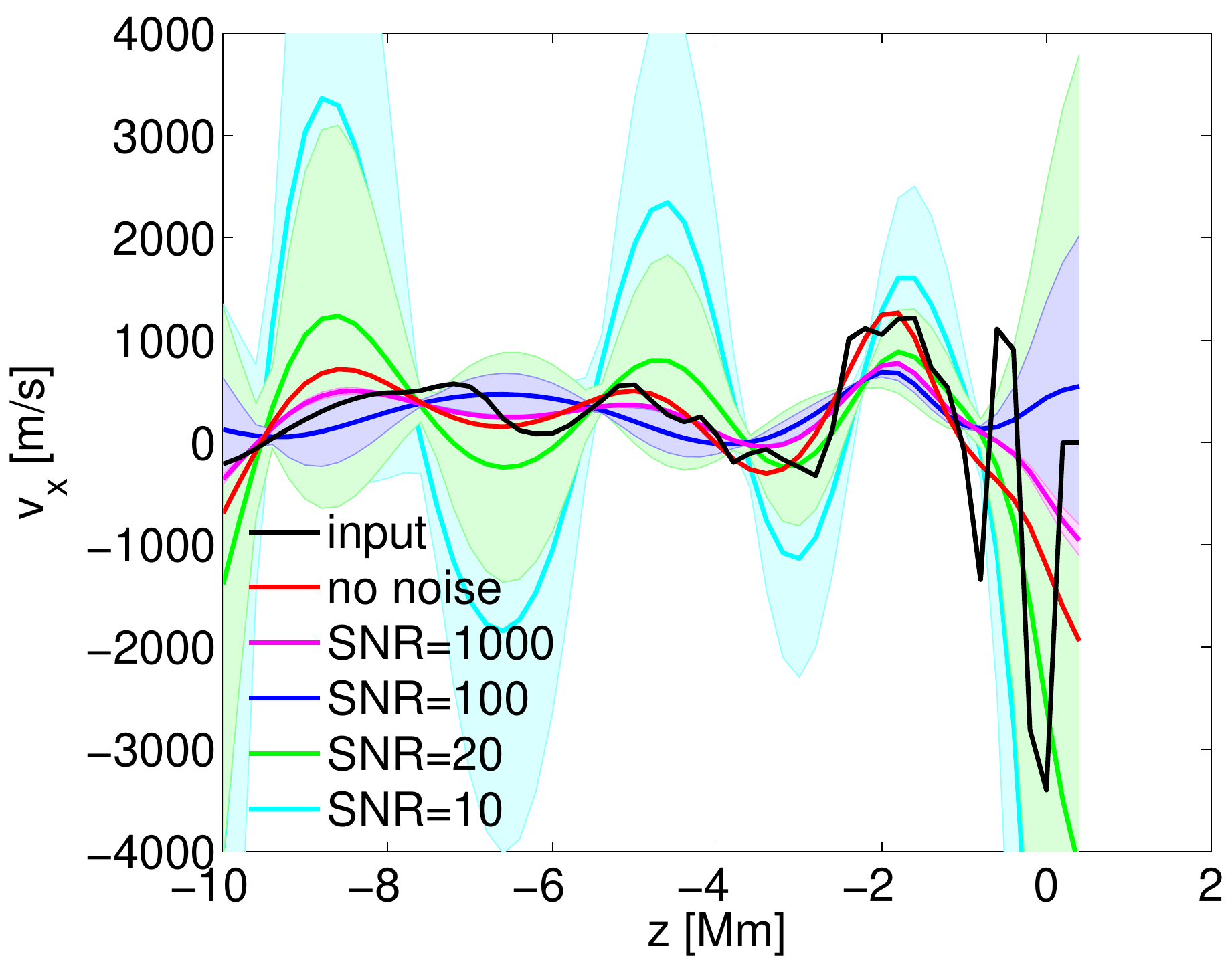}
\caption{Cut through the horizontal flow component at one point for the input velocity field (black) and reconstructions with varying noise levels (coloured lines). The shaded regions indicate the respective error bars. Note that the black and red lines do not have errors. }
\label{fig:comparison_cuts}
\end{figure}

{The reconstructed flow compared to the model is shown in} Figs.~\ref{fig:comparison}--\ref{fig:comparison_cuts}. One can see that the method works extremely well in case of no noise present. The correlation coefficient between Fig.~\ref{fig:comparison}a and \ref{fig:comparison}b is 0.8--0.9 for all components. At a first glance, almost all details of the flow field seem to be reconstructed correctly (see also first two rows of Fig.~\ref{fig:comparisonmaps}). Also the depth structure is reconstructed very well, which can be seen in Fig.~\ref{fig:comparison_cuts} (black and red lines). Small-scale oscillations of the flow are well beyond the resolution limit set by the averaging kernels, but the smoothed trend is captured quite well. Interestingly, the two cuts (black and red lines of Fig.~\ref{fig:comparison_cuts}) match better each other in the depth range of 0--6~Mm than deeper down. At depths 0--6~Mm all the averaging kernels overlap.

We did a further investigation of this issue and found that the best reconstruction is indeed achieved in depth range, which is ``included'' in several tomographic maps. This gives a recipe how to select the target function for SOLA inversions that may potentially lead to a reasonable estimate of the regularly gridded velocity field by our method. The target functions should be rather broad and they should overlap. Also a larger set of different target functions provide a more robust estimate of the true velocity field.  

Our reconstruction methods returns reasonable results also in the case when the random noise is present in the input data. The fact that the Wiener-like approach out-performs the ad-hoc regularisation is clear from Fig.~\ref{fig:comparisonmaps_methods}, where we compare example results of reconstruction methods differing by noise treatment. Obviously, without taking care of the noise, the results are completely wrong. The reconstructed flow magnitude oscillate from pixel to pixel, the noise is boosted by almost twelve orders. Adding an ad-hoc smoothing term improves the performance, however very large values of the smoothing parameter are needed to be chosen and even in such case the results are not satisfactory: They either still contain a large fraction of propagated random noise, or are very smooth with most of the details of the true velocity field lost. Wiener-like approach provides an optimal solution to the problem. Therefore we use this method further and study its properties.

A Wiener-like approach implemented by us allows a reasonable reconstruction of the flows for the signal-to-noise ratio (SNR\footnote{For instance, let us consider the inversion for supergranular flows. The typical magnitude of the horizontal flow components is roughly 300~\mps{}, reasonable inversions have the random-noise level of 25~\mps. In such a case SNR equals to 12. SNR of 100 would roughly represent the flow maps obtained after averaging of 100 supergranule individuals.}) larger than 20, positive correlation is found even in case of SNR of 10 (it means that the inverted flow maps are polluted with 10\% of random noise) -- see Table~\ref{tab:comparison}. 

\begin{table}
\caption{Correlation coefficient $\rho$ of the reconstructed flow components $v_x$, $v_y$, and $v_z$ with a Wiener-like dealing with random noise and the input synthetic velocity field for various noise levels. The correlation coefficient was computed from datacubes presented in Fig.~\ref{fig:comparison}.}
\label{tab:comparison}
\begin{tabular}{l|ccc}
SNR & $\rho_x$ & $\rho_y$ & $\rho_z$\\
\hline
\hline
\rule{0pt}{12pt}no noise & 0.87 & 0.81 & 0.91\\
1000 & 0.80 & 0.74 & 0.76\\
100 & 0.80 & 0.73 & 0.74\\
20 & 0.65 & 0.65 & 0.52\\
10 & 0.42 & 0.51 & 0.35
\end{tabular}

\end{table}

\begin{figure}
\includegraphics[width=0.45\textwidth]{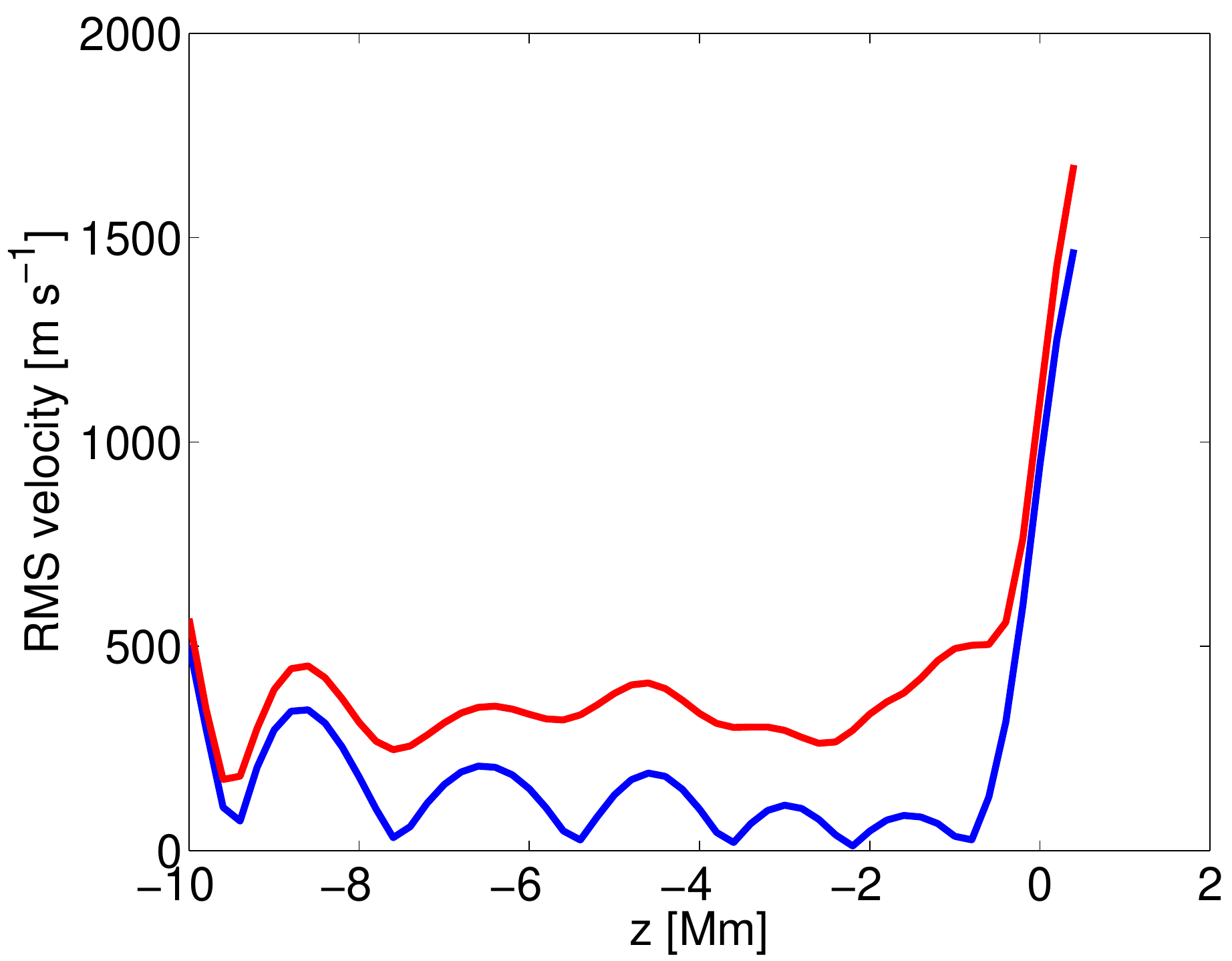}
\caption{Level of random noise in the reconstructed flow (blue) and the RMS of the reconstructed flow (red). Both curves are plotted for the horizontal component of velocity and for the case of SNR of 100. Obviously, the error increases beyond reasons at the ends of the domain.}
\label{fig:comparison_noise}
\end{figure}

The level of random noise is estimated following the procedure described in Section~\ref{sect:random_noise} for each depth. As an example we plotted the estimated RMS of the random noise in the reconstructed horizontal flow in Fig.~\ref{fig:comparison_noise} together with the RMS of the horizontal flow at the same depths for the case of SNR of 100. At all depths the SNR ratio of the reconstructed flow is above unity, at both ends of the domain SNR approaches unity. While at the bottom boundary this is due to the fact that the reconstruction is dominated by noise, at the top boundary the increase of both the noise RMS and the flow amplitude is the artifact of the normalisation by the new averaging kernel.  As seen in Fig.~\ref{fig:akerns}, at the top of the domain there is little power in all original averaging kernels $\cK$, hence the averaging kernels $\cF$ at these depths have an arbitrary shape with little localisation. As a consequence, the reconstruction of the flow at these depths is physically meaningless. 

It deserves to be noted that the reconstruction behaves reasonably in case the signal-to-noise ratio be large. Interestingly, this requirement conforms well with an conclusions described at the beginning of this Section, where we showed that rather broad averaging kernels lead to a better reconstruction of the regularly gridded estimate of the flow. In SOLA, localisation in the Sun (essentially the width of the averaging kernel) and the level of random noise are balanced by a trade-off parameter in the inversion cost function. Thus the choice of a broad target function automatically leads to a lower level of random noise and hence a larger SNR for a given tomographic map. The selection of a set of target functions that overlap prepares a set of tomographic maps suitable for the reconstruction of the regularly gridded estimate of a true velocity field. 

Even the Wiener-like solution is not ideal. The prior knowledge (or an estimate) of a spatial spectrum of the velocities and of the random noise poses a real challenge. It may be estimated reasonably perhaps for the surface for example in the way we did it in this paper, however it is expected from the theory that the spatial spectrum of the convective flows changes with depth. Such variability is very difficult to incorporate in the reconstruction. 

In the above described synthetic test we used the model averaging kernels (Fig.~\ref{fig:akerns}) which do not have a cross-talk components, whereas matrix $A$ [Eq. (\ref{eq:A})] allows to use the information even from the cross-talk (off-diagonal) terms. The presence of the cross-talk is an issue for inversions for weak quantities, such as the vertical component of the flow. In order to estimate the robustness of our method in the presence of cross-talk, we performed additional tests. First, the averaging kernels were computed in the same way as before, additionally we added cross-talk components that have the same shape as the components in the direction of the inversion. Such situation describes the case when the cross-talk is highly correlated with a wanted signal and its contribution is twice larger than the contribution of the flow component we invert for. Even in this case the reconstructed flow resembles the model one with a correlation coefficient of $\sim0.4$ for the horizontal components and $~0.2$ for the vertical component for the case of ${\rm SNR}=0.05$. Second, we kept only the cross-talk components and set the averaging kernels in the direction of the inversion to zero. The flow was recovered with a correlation coefficients $~0.5$ with the model one for both vertical and horizontal components. 

Additional tests showed that the correlation coefficient between the recovered flow and the model is lower when the cross-talk components of the averaging kernels are similar and comparable in peak magnitude to the averaging kernel in the direction of the inversion. A significant difference in peak magnitudes or an opposite sign actually help our method to improve the trustworthiness of the resulting reconstruction. The performed tests point out the difficulties caused by the presence of the cross-talks, however they also show on this badly posed examples that our reconstruction method is still able to recover the true structure of the flow reasonably.

\section{Summary}
We derived a method for deconvolution of a set of tomographic maps coming from local helioseismology, which allows to construct a regularly gridded estimate of the 3D vector velocity field. The method was successfully tested using synthetic data. We showed that it provides reasonable estimate of the true velocity field with a lesser smoothing than the original set of tomographic maps. To deal with a random noise, we introduce approach similar to Wiener filtering known from image processing. 

The method can naturally be applied to real data coming from SOLA inversions, which will be topic of our future study. It can be applied to any set of tomographic maps with a reasonable signal-to-noise ratio (larger than 50). The best usability can be expected in studies such as the investigation of the flow structure of solar supergranulation or similar features of convective flows. 

\begin{acknowledgements}
This work benefits from the Student grant awarded in 2014 to M.K. by Faculty of Mathematics and Physics, Charles University in Prague, which was supervised by M.\v{S}. M.\v{S}. acknowledges the support of the Czech Science Foundation (grant 14-04338S) and of the institute research project RVO:67985815 to Astronomical Institute of Czech Academy of Sciences. The help of Dominik Jan\v{c}\'\i{}k and Sven Ubik with the visualisation of the flows is greatly acknowledged. 
\end{acknowledgements}

\bibliographystyle{aa}
\bibliography{inversions}
\begin{appendix}

\section{Derivation of Equation (\ref{eq:oneline})}
\label{app:oneline}
In Section~\ref{sect:theproblem} we prescribed ad-hoc a set of Equations~(\ref{eq:oneline}) to be solved to obtain an estimate of the reconstructed 3D vector flow defined on a regular grid. In the following we will show that these equations may be properly obtained by minimisation procedure. 

We define a misfit between the inverted flow map at target depth $z^d$ and the modelled tomographic map as
\begin{align}
\chi^2_{z^d}(\bvec{r})=&\sum\limits_\alpha \left[ \vinv_\alpha(\bvec{r};z^d)-h_x^2 h_z \sum\limits_{\beta,{\mathbf r}_j,z} \cK_\beta^\alpha(\bvec{r}_j-\bvec{r},z;z^d)v_\beta(\bvec{r}_j,z)\right]^2\times\nonumber\\
&\times \frac{1}{(\sigma^d_\alpha)^2},
\label{eq:misfit}
\end{align}

For each target depth $z^d$ and positional vector $\bvec{r}$ there is a separate Eq.~(\ref{eq:misfit}), however the modelled velocity field $\bvec{v}$ is common to all these equations. Note the inverse  weighting by the RMS of the random noise discussed already in Section~\ref{sect:theproblem}.

Equation~(\ref{eq:misfit}) can be seen as a cost function, which we want to minimise with respect to the unknown velocity field $\bvec{v}$. Using standard methods of calculus we first obtain equation
\begin{align}
0=&\frac{\partial \chi^2_{z^d}(\bvec{r})}{\partial v_\gamma(\bvec{r_1},z_1)} =   \sum_{\alpha} \frac{2}{\left(\sigma^d_\alpha\right)^2}  \left(-h_x^2h_z\cK^\alpha_\gamma(\bvec{r}_1-\bvec{r},z_1,z^d)\right) \times \nonumber\\
& \times \left[v_\alpha^{\rm inv}(\bvec{r};z^d) - h_x^2h_z\sum_\beta\sum\limits_{{\mathbf r}_j}\sum\limits_z \cK_\beta^\alpha(\bvec{r}_j-\bvec{r},z;z^d) v_\beta(\bvec{r}_j,z)\right],
\end{align}
which holds for each $z^d$ and $\gamma$. The minimisation is performed with respect to the $\gamma$-component of velocity at point $(\bvec{r}_1,z_1)$. Since in each of these points the equation equals to zero, we may sum over all these points in the computational domain. For each $\gamma$ we obtain
\begin{align} 
0=& \sum_{{\mathbf r}_1}\sum_{z_1} \sum_{\alpha} \frac{2}{\left(\sigma^d_\alpha\right)^2}  \left(-h_x^2h_z\cK^\alpha_\gamma(\bvec{r}_1-\bvec{r},z_1,z^d)\right) \times \nonumber\\
& \times \left[v_\alpha^{\rm inv}(\bvec{r};z^d) - h_x^2h_z\sum_\beta\sum\limits_{{\mathbf r}_j}\sum\limits_z \cK_\beta^\alpha(\bvec{r}_j-\bvec{r},z;z^d) v_\beta(\bvec{r}_j,z)\right].
\label{eq:real}
\end{align}
Note that the first term reminds the normalisation condition to the averaging kernels 
\begin{equation}
\sum_{z_1}\sum_{{\mathbf r}_1} h_x^2 h_z \cK_\gamma^\alpha(\bvec{r}_1,z_1;z^d)=\delta^\alpha_\gamma,
\end{equation}
where $\delta^\alpha_\gamma$ is a Kronecker delta. We solve (\ref{eq:real}) in the Fourier space using a definition given by a pair of Equations (\ref{eq:Ffor}) and (\ref{eq:Finv}) to obtain
\begin{align}
0=&-\frac{1}{\left(\sigma^d_\gamma\right)^2}h_k^2\sum_{\mathbf k} \tilde v_\gamma^{\rm inv}(\bvec{k}; z^d)\exp{\left[ \ii\bvec{k} \cdot \bvec{r}\right]} + \nonumber\\
&+\frac{1}{\left(\sigma^d_\gamma\right)^2} h_x^2 h_z h_k^4 \sum_\beta\sum_{{\mathbf r}_j}\sum_z\sum_{\mathbf k_1}  \tilde{\cK}^\gamma_\beta(\bvec{k}_1,z;z^d) \times \nonumber\\
&\times\exp{\left[\ii\bvec{k}_1 \cdot \bvec{r}_j\right]} \exp{\left[-\ii\bvec{k}_1 \cdot \bvec{r}\right]} \sum_{\mathbf k_2}\tilde v_\beta(\bvec{k}_2,z)\exp{\left[\ii\bvec{k}_2 \cdot \bvec{r}_j\right]}.
\label{eq:eachr}
\end{align}
Equation (\ref{eq:eachr}) holds for each $z^d$, each $\gamma$, and each $\bvec{r}$. In the following we use the orthogonality of the functions $\exp\left[ \ii\bvec{k} \cdot \bvec{r}\right]$ and solve (\ref{eq:eachr}) in the $\bvec{k}$-by-$\bvec{k}$ manner. After some algebra we obtain
\begin{equation}
\frac{1}{\left(\sigma^d_\gamma\right)^2} \tilde v_\gamma^{\rm inv}(\bvec{k};z^d) = \frac{(2\pi)^2}{\left(\sigma^d_\gamma\right)^2}   h_z  \sum_\beta\sum_z \tilde{\cK}^{\gamma*}_\beta(\bvec{k},z;z^d) \tilde v_\beta(\bvec{k},z),\quad \forall \bvec{k},
\end {equation}
which is the Equation~(\ref{eq:oneline}).

\section{Estimates of spatial power spectra}
\label{app:spectra}
In SOLA, the cost function \citep[Equation~(12) in ][]{Svanda2011} allows to balance between the localisation (the averaging kernel) and the random-noise level in the resulting flow map. Symbolically written
\begin{equation}
{\rm cost\ function} = {\rm avg.\ kernel\ misfit} + \mu\, {\rm random\ noise},
\label{eq:solacost}
\end{equation}
where ${\rm misfit}$ evaluates the localisation of the averaging kernel and $\mu$ is a trade-off parameter free of choice. For details we refer the reader to \cite{1992AA...262L..33P}. A minimisation of this cost function gives at the end of the whole procedure a flow map at a given depth together with a averaging kernel and an estimate for a level of random noise. By choosing values of the trade-off parameter $\mu$ one can easily influence the random-noise level in these maps. 

From the resulting flow map one can compute a spatial power spectrum $P(\bvec{k})$ using standard methods. Let us consider two extreme cases. First let us choose $\mu$ large, so that the random-noise term multiplied by $\mu$ dominates the right-hand side of inversion cost function (\ref{eq:solacost}). This selection causes the random-noise term to be preferentially minimised and the resulting flow maps contain mostly the signal. The resulting spatial power spectrum $P(\bvec{k})$ approximates the power spectrum of the flows, let us name it $S(\bvec{k})$. The drawback of this selection usually is a worse localisation in the Sun (a worse-quality averaging kernels with sidelobes).  

As a second extreme, let us select $\mu$ small. Then the misfit term in the cost function (\ref{eq:solacost}) is preferred and the resulting flow map contains a large fraction of noise. It even might be noise dominated, so that the level of random noise is much larger (by many orders) than the expected magnitude of the real flows. The power spectrum $P(\bvec{k})$ approximates the power spectrum of the noise, nicked as $N(\bvec{k})$. The localisation in the Sun is usually good in this case. 

The estimates for the power spectrum of the flows $S$ and the noise $N$ would not be robust enough if only inversion for a single flow maps would be considered. Also in case of a single map, the difference between the averaging kernels for the two kinds of inversions could be large, so that the validity of our approach to obtain the power-spectra estimates might be questioned. To improve the robustness one need to average both power spectra over a larger set of flow maps. We used 1100 individual flow maps obtained with both signal- and noise-dominated inversions to estimate power spectra $S$ and $N$. Averaging over large sample of flow maps also allows to tune values of needed $\mu$ so that the difference (evaluated by eye) of the resulting averaging kernels for noise- and signal-dominated inversion is not too large. 

We believe that the described methodology allows to derive robust estimates for spatial power spectra of the flows $S$ and the noise $N$. Both power spectra might then be easily scaled so that their RMS values attain desired values by simply invoking the Parseval theorem and the linearity between the RMS and the scaling of the random sample. 
 
\begin{figure}[h]
\includegraphics[width=0.49\textwidth]{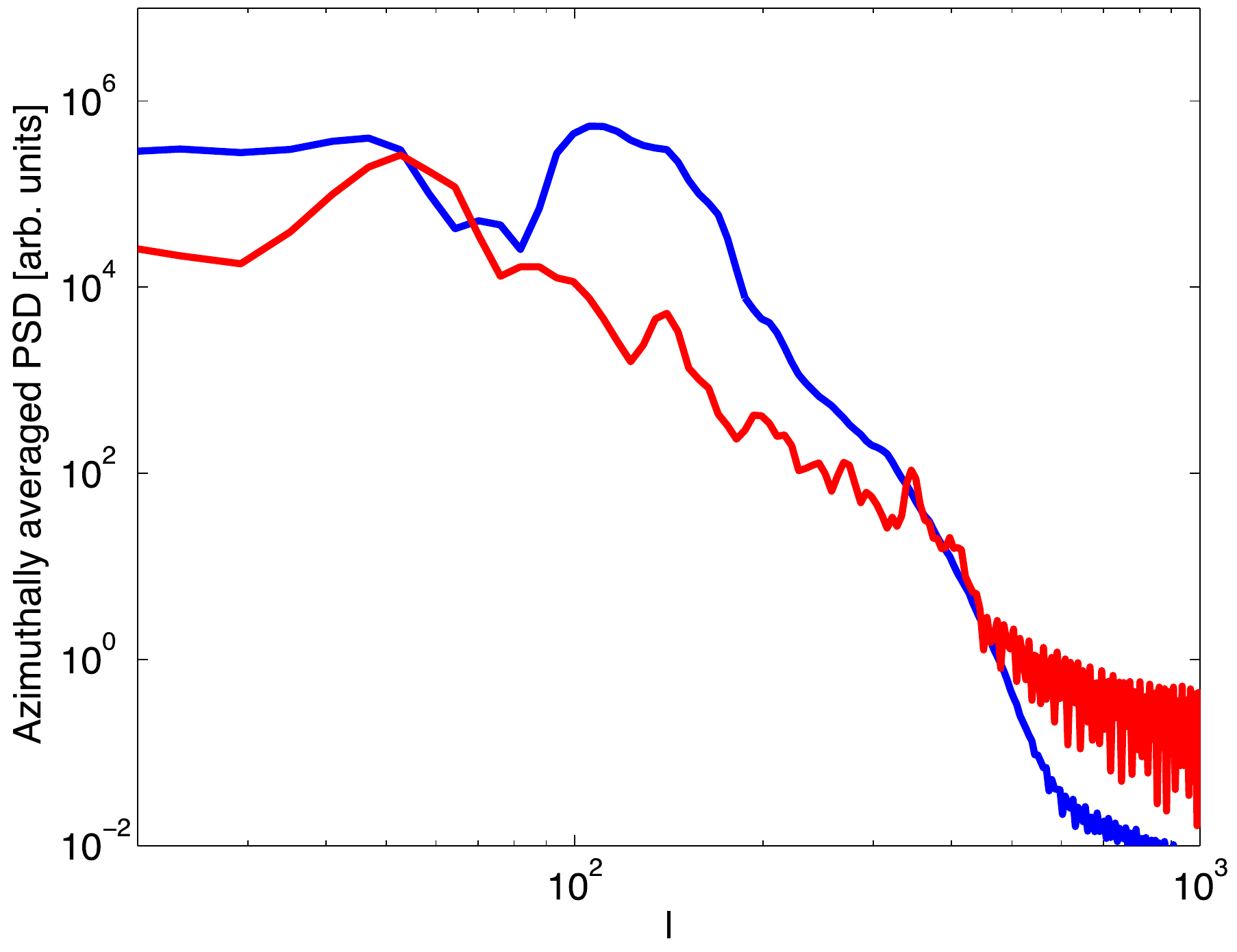}
\caption{Estimates of power spectra of the signal (blue) and the noise (red) obtained from the real data as a function of angular degree $l$. The power spectrum of the signal (flows) were normalised so that the RMS of the flow is 100~\mps. The power spectrum of the random noise was also normalised to RMS of 10~\mps. Such situation therefore represents ${\rm SNR}=10$. }
\label{fig:powerspectra}
\end{figure}

Examples can be seen in Fig.~\ref{fig:powerspectra} for a SNR of 10. In such situation, obviously the noise dominates the signal at large angular degrees, the situation is largely reversed at supergranular scales at $l\sim 120$ and then again in the low-$l$ part of the spectrum. 

\end{appendix}
\end{document}